# Magnetic Field Structure in Spheroidal Star-Forming Clouds. II. Estimating Field Structure from Observed Maps

*short title:* Magnetic Field Structure in Spheroids. II.

Philip C. Myers
Center for Astrophysics | Harvard and Smithsonian (CfA), Cambridge, MA 02138, USA
pmyers@cfa.harvard.edu

Ian W. Stephens
Center for Astrophysics | Harvard and Smithsonian (CfA), Cambridge, MA 02138, USA

Sayantan Auddy
Academia Sinica Institute for Astronomy and Astrophysics (ASIAA), Taipei 106, Taiwan

Shantanu Basu
University of Western Ontario (UWO), London, Ontario N6A 3K7, Canada

Tyler L. Bourke
SKA Organisation, Jodrell Bank Observatory, Lower Withington, Macclesfield SK11 9DL, UK

and

Charles L. H. Hull[1]
National Astronomical Observatory of Japan, NAOJ Chile, Alonso de Cordova 3788, Office 61B, 7630422, Vitacura, Santiago, Chile
Joint ALMA Observatory, Alonso de Cordova 3107, Vitacura, Santiago, Chile
[1]NAOJ Fellow

**Abstract**

This paper presents models to estimate the structure of density and magnetic field strength in spheroidal condensations, from maps of their column density and their polarization of magnetically aligned dust grains. The density model is obtained by fitting a column density map with an embedded $p = 2$ Plummer spheroid of any aspect ratio and inclination. The magnetic properties are based on the density model, on the Davis-Chandrasekhar-Fermi (DCF) model of Alfvénic fluctuations, and on the Spheroid Flux Freezing (SFF) model of mass and flux conservation in Paper I. The field strength model has the resolution of the column density map, which is finer than the resolution of the DCF estimate of field strength. The models are applied to ALMA observations of the envelope of the protostar BHR71 IRS1. Column density fits give the density model, from $(2.0 \pm 0.4) \times 10^5$ cm$^{-3}$ to $(7 \pm 1) \times 10^7$ cm$^{-3}$. The density model predicts the field directions map, which fits the polarization map best within 1100 au, with standard deviation of angle differences 17°. In this region the DCF mean field strength is $0.7 \pm 0.2$ mG and the envelope mass is supercritical, with ratio of mass to magnetic critical mass $1.5 \pm 0.4$. The SFF field strength profile scales with the DCF field strength, from $60 \pm 10$ μG to $4 \pm 1$ mG. The spatial resolution of the SFF field strength estimate is finer than the DCF resolution by a factor ~7, and the peak SFF field strength exceeds the DCF field strength by a factor ~5.

# 1. INTRODUCTION

Magnetic fields and their structure are important features of the evolution of star-forming clouds, but their exact role remains uncertain. In many star-forming regions, magnetic forces are believed to be weaker than gravitational forces according to Zeeman observations (Crutcher 2012) and numerical simulations (Chen & Ostriker 2015). If so, they cannot prevent star-forming collapse and they cannot set the mass scale of the IMF (Krumholz & Federrath 2019). Nonetheless, in some star-forming regions the mass-to-flux ratio is only slightly supercritical, and the field energy exceeds the turbulent energy (e.g., NGC1333 IRAS4A, Frau et al. 2011; G31.41+0.31, Beltrán et al. 2019). When magnetic fields are energetically significant they oppose or channel turbulent flows, and they drive outflows, which reduce global star formation rates by factors of a few (Federrath et al. 2014).

The relative importance of magnetic and gravitational energy may be addressed by studying the variation of field strength $B$ with density $n$, since the field is expected to vary as $B \sim n^{\kappa}$ with $\kappa =$ 2/3 for fields weak compared to gravity, but with $\kappa$ closer to ½ for fields strong compared to gravity (Crutcher 2012, Tritsis et al. 2015, Mocz et al. 2017, Kandori et al. 2018). The relative importance of magnetic and turbulent energy can be studied by observations of polarization due to magnetically aligned grains. The most widely used method of estimating field strength from polarization patterns attributes random fluctuations in polarization direction to Alfvénic fluctuations in field strength, after removal of an ordered component (Davis 1951; Chandrasekhar & Fermi 1953; hereafter DCF).

Polarimetric observations sensitive to magnetic field structure have been made at submillimeter wavelengths with telescopes including ALMA, the SMA, the JCMT, and SOFIA, and at near-infrared wavelengths with numerous telescopes (Pattle & Fissel 2019). These observations show highly ordered "hourglass" structure in some regions, but in most regions the patterns have a different order, or they are more chaotic (Hull et al. 2017, Hull & Zhang 2019).

Analysis of these observations requires comparison with models of field structure to assess the structure and role of the magnetic field. The most complex models come from numerical simulations of magnetized clouds (e.g., Teyssier & Commercon 2019, Hennebelle & Inutsuka 2019). However it is difficult to match such simulations to observations in detail, due to the large number of model parameters and to the relatively long run times of simulations.

In contrast, an analytic model of the magnetic field structure of a condensation can more easily generate model maps to compare with observed polarization maps (Tomisaka 2011, Kataoka et al. 2012, Padovani et al. 2012, Reissl et al. 2014). When the magnetic field structure is directly related to the density structure, a density model can predict maps of both magnetic field direction and column density, for comparison with observed maps.

Such column density maps can give information on a scale finer than the resolution of the DCF field strength. The DCF estimate requires $N \gg 1$ independent polarization measurements for a statistically significant estimate of polarization angle dispersion. In the usual case where the polarization and column density maps have the same resolution, the column density map has a resolution advantage of a factor $\sim N^{1/2}$ over the DCF field strength estimate. In the example of BHR71 IRS1 analyzed in Section 5, this resolution advantage is a factor of ~7.

Analytic models of this finer-scale structure cannot be applied in all cases. The associated column density map may be too poorly defined, due to embedded protostars or uncertain temperature structure (Frau et al. 2011, Alves et al. 2018). The geometry of the density model may not always match the map, as when the model symmetry is spherical (e.g. Mestel 1966, hereafter M66) or toroidal (Li & Shu 1996). These simple shapes cannot match the map contours of a significant number of star-forming regions, which may resemble elongated filaments (e.g. Arzoumanian et al. 2011) or flattened circumstellar envelopes (e.g. Chiang et al. 2010). Nonetheless, when a polarization map is associated with a suitable column density map, a matching analytic model offers a higher-resolution estimate of field structure than is possible with DCF analysis alone.

This paper presents an analytic model which can be used to infer magnetic properties of observed condensations with a range of shapes, by predicting model maps of both field direction and column density. The models in Paper I and in this paper extend the pioneering model of M66. In M66, a spherical region of uniform original density and field strength contracts into a centrally condensed sphere while conserving mass and flux. The inward advection of field lines during contraction causes the magnetic field lines to develop an hourglass shape. This "primary distortion" field structure is calculated as a first approximation, assuming that the field is too weak to prevent radial contraction.

This paper does not treat the "secondary distortion" of M66, which develops from the unbalanced magnetic force toward the midplane, and which concentrates inflowing gas in an equatorial "pseudodisk" (Galli & Shu 1993a,b). Similarly, this work does not assume that the field line configuration is due to force balance between self-gravity and field line tension, or between self-gravity and magnetic pressure gradient.

The spherically symmetric model of M66 was extended to prolate and oblate spheroids, by assuming that the spheroid maintains its shape during contraction. The basic properties of this "Spheroid Flux Freezing" (SFF) model were described in Myers et al. (2018), hereafter Paper I. In the simplest implementation, field directions in the plane of the sky are matched to field directions inferred from observed polarization directions. In a more accurate implementation, the field directions are integrated along each line of sight to simulate the expected polarization directions, as in the studies of Tomisaka (2011), Kataoka et al. (2012), Padovani et al. (2012), Reissl et al. (2014).

Paper I showed that poloidal flux tube patterns and density contours in the plane of the sky can be obtained from simple analytic expressions for Plummer spheroids of given density contrast, aspect ratio and inclination angle. These patterns agree with numerical models and with observed polarization maps, within uncertainties similar to those in recent observational studies. The methods in Paper I have been used to analyze polarization observations of the protostellar envelope VLA1623A (Sadavoy et al. 2018), the starless core FeSt 1-457 (Kandori et al. 2020a), and the cluster-forming region Serpens South (Pillai et al. 2020).

This paper extends these results to allow estimates of the structure of magnetic field direction and strength, the mass-to-critical-mass ratio, and the peak field strength in centrally condensed regions of spherical, oblate, and prolate shape. Section 2 gives new expressions for maps of spheroid column density and field direction, for four spheroid types having any aspect ratio and inclination. These expressions include field directions for an oblate spheroid with a significant toroidal field component. Section 2 also gives an improved model of the background density and column density. Section 3 extends the DCF method to obtain the inclination-averaged field strength for a given level of Alfvén wave excitation, or equivalently for the dispersion in the angle difference between polarization and model. This mean field strength is combined with the SFF density model to obtain the ratio of mass to critical mass, and to obtain the profile of field strength as a function of equatorial radius. Section 4 summarizes step-by-step application of the models in Sections 2 and 3. Section 5 describes field strength estimates from ALMA maps of the envelope of the protostar BHR71 IRS1. Section 6 discusses the results and Section 7 summarizes the conclusions.

## 2. DENSITY, COLUMN DENSITY, AND FIELD DIRECTION MODELS

This section and Section 3 present the analytical basis of the magnetic field estimation. These sections define the spheroid density model, and they give equations for the column density models depending on spheroid orientation. Expressions for magnetic field direction are derived from the density model and from flux freezing and mass conservation. The structure of the magnetic field strength is derived from the mean field strength and from the scaling of the field strength with density. Readers primarily interested in results may prefer to concentrate on Sections 4-7, with guidance from Tables 1-2 in this section, from the model geometry in Figure 9 in Appendix A, and from the list of symbols in Table 3 in Appendix B.

## 2.1. Spheroid Geometry

In the adopted coordinate system, $y$ increases away from the observer along the line of sight, and the $x$- and $z$-axes lie in the plane of the sky. The polarization and column density maps are each centered on the position $(x, z) = (0,0)$ of the column density map peak. The $x$-$z$ plane is oriented so that the $z$-axis coincides with the symmetry axis of the observed polarization structure, and thus with the projection of the magnetic axis on the plane of the sky. The angle between celestial north (N) and the z-axis is denoted $\theta_M$, increasing clockwise from N.

In the $x$-$y$-$z$ coordinate system, a given radius vector $r$ has inclination angle $i$ increasing from the $z$-axis toward $r$ in the $r$-$z$ plane. The projection of $r$ onto the $x$-$z$ plane lies at polar angle $\theta$, increasing from the $z$-axis toward the $x$-axis. The projection of $r$ onto the $x$-$y$ plane lies at azimuth angle $\phi = \cos^{-1}(\tan\theta/\tan i)$ for $|\tan\theta/\tan i| \leq 1$. These properties are illustrated in Figure 9 in Appendix A.

The $p = 2$ Plummer sphere and four types of $p = 2$ Plummer spheroid are considered here. The symmetry axis of each spheroid is assumed to lie either parallel or perpendicular to the magnetic axis, as illustrated in Tables 1 and 2. When the magnetic and symmetry axes are coincident, the spheroid is called "parallel prolate" or "parallel oblate." Their common axis direction has inclination angle $i$ from the z-axis toward the observer in the $y$-$z$ plane, in the direction $\hat{z}\cos i - \hat{y}\sin i$. When the magnetic and symmetry axes are perpendicular, the spheroid is called "perpendicular prolate" or "perpendicular oblate." Then the magnetic axis direction is again $\hat{z}\cos i - \hat{y}\sin i$ but the symmetry axis is along the $x$-direction. Spheroids whose magnetic and symmetry axes are neither parallel nor perpendicular are outside the scope of this model.

When the spheroid is perpendicular prolate or perpendicular oblate, its projected symmetry axis lies along the x-axis, but its true symmetry axis may lie at an angle $\phi \neq 0$ from the $x$-axis. Its length in the $x$ - direction is then less than its true length for the perpendicular prolate spheroid, and greater than the true length for the perpendicular oblate spheroid. At present it does not appear possible to infer $\phi$ directly from the polarization and column density maps. Thus it is assumed for the perpendicular prolate spheroid and the perpendicular oblate spheroid that the symmetry axis lies in the plane of the sky, i.e. that $\phi = 0$.

## 2.2. Cartesian Coordinate Relations

The model coordinate origin $(x, z) = (0,0)$ is chosen to coincide with the column density map peak, which is here denoted in celestial coordinates as $(x_c, z_c) = (x_{c0}, z_{c0})$. Here the celestial coordinate $x_c$ is chosen to increase with decreasing Right Ascension (to the W) and $z_c$ increases with increasing declination (to the N). The symmetry axis of the prolate or oblate perpendicular spheroid lies along the $x$ – axis, as assumed above in Section 2.1. This axis lies at angle $\theta_M$ from the E-W direction as shown in Figure 9. In contrast, the symmetry axis of the prolate or oblate parallel spheroid lies in the $y$-$z$ plane, inclined through angle $i$ from the $z$- axis. It lies in the plane of the sky only when $i = 0$. Its projection onto the plane onto the sky lies at angle $\theta_M$ from the N-S direction. With these definitions, the model and celestial coordinates of a point in the plane of the sky are related by

$$x = (x_c - x_{c0}) \cos \theta_M + (z_c - z_{c0}) \sin \theta_M \qquad (1)$$

and

$$z = (x_c - x_{c0}) \sin \theta_M + (z_c - z_{c0}) \cos \theta_M \quad . \qquad (2)$$

Figure 9 illustrates the case where the center of the celestial map coincides with the center of the model map, i.e. when $x_{c0} = 0$ and $z_{c0} = 0$.

## 2.3. Density

The density model is chosen to resemble centrally condensed star-forming regions which have a single local maximum, and whose shape may be spherical (e.g., Tafalla et al. 2004), elongated (e.g., Arzoumanian et al. 2011), or flattened (e.g., Sadavoy et al. 2018). Here "density" refers to number density or volume density, as opposed to mass density or column density. The adopted density model is a $p = 2$ Plummer spheroid (Plummer 1911) centered at $x = y = z = 0$ and embedded in a uniform background medium of density $n_u$, as in Paper I,

$$\nu = 1 + \nu_0 (1 + \omega^2)^{-1} \, . \qquad (3)$$

The uniform background medium is an essential part of the flux-freezing model, as discussed in Section 2.6.

In equation (3) $\nu \equiv n/n_u$ is the density normalized by the uniform background density $n_u$, and $1 + \nu_0$ is the normalized peak density, where $\nu_0 \equiv n_0/n_u$ is the density contrast ratio. The density structure can be considered as a system of concentric spheroidal surfaces, each surface having normalized radius $\omega$ and normalized density $\nu$, where $\nu$ decreases with increasing $\omega$.

When the spheroid is a sphere, $\omega^2$ is given by $\omega^2 = \xi^2 + \eta^2 + \zeta^2 = (x/r_0)^2 + (y/r_0)^2 + (z/r_0)^2$. Each dimensionless coordinate is equal to its space coordinate, normalized by the scale length $r_0$. Here $r_0$ is the shortest distance from the origin to the spheroidal surface whose normalized density is $\nu = 1 + \nu_0/2$. When the spheroid is prolate or oblate, the aspect ratio $A > 1$ is the ratio of its greatest and smallest principal-axis diameters. When the spheroid symmetry axis coincides with the $x$-axis, the prolate spheroid has $\omega^2 = (\xi/A)^2 + \eta^2 + \zeta^2$ while the oblate spheroid has $\omega^2 = \xi^2 + (\eta/A)^2 + (\zeta/A)^2$. When the spheroid symmetry axis coincides with the $z$-axis, the prolate spheroid has $\omega^2 = \xi^2 + \eta^2 + (\zeta/A)^2$, while the oblate spheroid has $\omega^2 = (\xi/A)^2 + (\eta/A)^2 + \zeta^2$. These relations indicate that a spheroidal surface of constant $\omega$ has smallest radius $\omega r_0$ in the short-axis direction and greatest radius $A\omega r_0$ in the long-axis direction.

Integration of $\nu$ over $0 \leq \omega' \leq \omega$ gives the normalized mean density $\bar{\nu}(\omega)$,

$$\bar{\nu} = 1 + (3\nu_0 \omega^{-2})(1 - \omega^{-1} \tan^{-1} \omega) \qquad . \qquad (4)$$

The ratio of density to mean density is $t \equiv \nu/\bar{\nu}$, or

$$t = \frac{1+\nu_0(1+\omega^2)^{-1}}{1+(3\nu_0\omega^{-2})(1-\omega^{-1}\tan^{-1}\omega)} \qquad . \qquad (5)$$

Here $t$ approaches its maximum value of 1 when $\omega \to 0$ or when $\omega \to \infty$. Otherwise $t < 1$, e.g. $t = 0.5$ when $\nu_0 = 100$ and $\omega = 5$. Equation (5) is used to calculate magnetic field directions in the plane of the sky. It provides $t_q$ for substitution into equation (23) along with values of $\omega^2$ given in Table 2, as discussed in Section 2.8.

Equations (3) - (5) for density-related properties apply to spheroids of spherical, oblate, or prolate shape. They depend on spheroid shape only through the differing dependence of $\omega$ on $A$ as given above. In contrast, the mass of a spheroid with properties $\nu_0$, $\omega$, and $A$ also depends on an additional factor, which is an integer power of $A$. Then the mass within $\omega$ is $M(\omega) = (4\pi/3)[(\omega r_0)^3] m n_u \bar{\nu}(\omega) A^k$, where $m$ is the mean molecular mass and where $k = 0$ for the sphere, $k = 1$ for the prolate spheroid, and $k = 2$ for the oblate spheroid.

### 2.4. Column Density

The column density is obtained by integrating the density in equation (3) along the line of sight, i.e. along a line parallel to the $y$- axis. For the simplest background model, $2Y$ is the extent of the original medium of density $n_u$, where $Y$ is much greater than the plane-of-the-sky extent of the observed column density map. Then the column density is given by $N(\xi, \zeta) = N_u + \Delta N(\xi, \zeta)$, where the background column density is $N_u \equiv 2 n_u Y$ and where the background-subtracted column density is

$$\Delta N(\xi, \zeta) \equiv 2 n_0 r_0 \int_0^{Y/r_0} d\eta\, \{1 + [\omega(\xi, \eta, \zeta)]^2\}^{-1} . \qquad (6)$$

Evaluation of equation (6) yields a factor of form $\tan^{-1}[(Y/r_0)(1 + \omega^2)^{-1/2}]$. It is assumed that $Y$ is great enough to approximate this factor by $\tan^{-1} \infty = \pi/2$. A more detailed background model is described in Section 2.6.

For each the four spheroid types considered here, $\omega^2$ has a different dependence on $\eta$, so evaluation of equation (6) gives four expressions for $\Delta N$. These are $\Delta N_{\|p}$ for the parallel prolate spheroid, $\Delta N_{\perp p}$ for the perpendicular prolate spheroid, $\Delta N_{\|o}$ for the parallel oblate spheroid, and $\Delta N_{\perp o}$ for the perpendicular oblate spheroid:

$$\Delta N_{\|p}(\xi, \zeta) = \frac{\pi n_0 r_0 A}{\{(1+\xi^2)[1+(A^2-1)(\cos i)^2]+\zeta^2\}^{1/2}} \qquad (7)$$

$$\Delta N_{\perp p}(\xi, \zeta) = \frac{\pi n_0 r_0}{[1+(\xi/A)^2+\zeta^2]^{1/2}} \qquad (8)$$

$$\Delta N_{\|o}(\xi, \zeta) = \frac{\pi n_0 r_0 A}{\{(1+(\xi/A)^2)[1+(A^2-1)(\sin i)^2]+\zeta^2\}^{1/2}} , \qquad (9)$$

and

$$\Delta N_{\perp o}(\xi, \zeta) = \frac{\pi n_0 r_0 A}{[1+\xi^2+(\zeta/A)^2]^{1/2}} \quad . \quad (10)$$

For the parallel spheroids in equations (7) and (9), $\Delta N$ depends on inclination $i$. In contrast, $\Delta N$ is independent of $i$ for the perpendicular spheroids in equations (8) and (10), because they have rotational symmetry about the $x$-axis. For convenience Table 2 summarizes equations (7) - (10) along with cartoons of the perpendicular spheroids and the uninclined parallel spheroids. When $A = 1$ each equation reduces to that for a $p = 2$ Plummer sphere, $\Delta N_s = \pi n_0 r_0 (1 + \xi^2 + \zeta^2)^{-1/2}$, as expected.

### 2.5. Column Density Parameters

To obtain the best possible estimate of an observed source structure, the appropriate model equation should be fit to the observed map. Initial estimates for the aspect ratio $A$, the scale length $r_0$, and the peak density $n_0$ can be made when the column density map is rendered in contours of constant $\Delta N$. When the spheroidal model matches the map sufficiently well, each column density contour approximates the shape of an ellipse. Then for a given $\Delta N$, the peak column density $\Delta N_{\max}$ and the ellipse principal axis radii $x_0$ and $z_0$ provide initial guesses for $A$, $r_0$, and $n_0$, based on equations (7) - (10). Table 1 gives expressions for these quantities for the perpendicular spheroids, which do not depend on inclination $i$, and for the parallel spheroids, which require an assumed value for $i$.

### 2.6. Background Density and Column Density

The background medium is a necessary part of the density and field direction models, because these models depend on mass and flux conservation during contraction from an original medium (M66). This section derives the background density from a column density map, and from assumed ratios of column density and mass between the condensation and its original medium.

The condensation and its original medium are each assumed to have spheroidal symmetry, with the same center, aspect ratio, and orientation. The original medium has uniform density and field strength, and its bounding spheroidal surface has finite spatial extent. This finite boundary is more physical and realistic than the infinite original medium assumed by M66 and Paper 1. The spheroid shape is assumed for simplicity to be perpendicular prolate or perpendicular oblate since the column density of these spheroids does not change with inclination.

The lowest closed column density contour of the condensation map has short-axis radius $b_l$ and column density $N_l$, and this contour encloses mass $M_l$ with normalized mean density $\bar{v}_l$. The original spheroid has uniform density $n_u$, short-axis radius $b_u$, mean column density $N_u$, and enclosed mass $M_u$. Here the subscript "$l$" refers to the lowest closed contour of the column density map, and the subscript "u" refers to the original region of uniform density and field strength. Let $\epsilon_l \equiv M_l/M_u$ be the "formation efficiency" or mass ratio of the condensation to its original spheroid, and let $v_{Nl} \equiv N_l/N_u$ be the column density ratio of the lowest-contour condensation to its original spheroid. Then conservation of mass implies that the dimensions of the condensation and the original spheroid are related by $b_l/b_u = (\epsilon_l/\bar{v}_l)^{1/3}$.

The column density $N_l$ can be written following equations (8) and (10) as

$$N_l = N_u + \frac{\pi n_0 r_0 A^{k-1}}{(1+\omega_l^2)^{1/2}} \qquad (11)$$

where $A$ is the spheroid aspect ratio, with $A = 1$ for the sphere. The aspherical spheroids have $k = 1$ if prolate and $k = 2$ if oblate. The background column density is assumed to equal the mean column density of the original medium,

$$N_u = (4/3)n_u b_l (\bar{v}_l/\epsilon_l)^{1/3} A^{k-1} \ . \qquad (12)$$

Combining equations (11) and (12) with the relations $v_0 = n_0/n_u$ and $\omega_l = b_l/r_0$ gives the column density ratio $v_{Nl} = N_l/N_u$,

$$v_{Nl} = 1 + \frac{3\pi v_0 (\epsilon_l/\bar{v}_l)^{1/3}}{4\omega_l(1+\omega_l^2)^{1/2}} \ , \qquad (13)$$

which is independent of spheroid type and aspect ratio.

Table 1

Estimation of $p = 2$ Plummer spheroid parameters from column density map contours

| Spheroid type | contour of $\nu \equiv \Delta N/\Delta N_{max}$ | aspect ratio $A$ | scale length $r_0$ | peak density $n_0$ |
|---|---|---|---|---|
| ∥ prolate ($i = 0$ view) 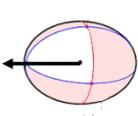 | 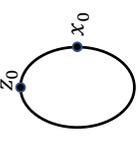 | $A = \left[1 + \dfrac{(z_0/x_0)^2 - 1}{(\cos i)^2}\right]^{1/2}$ | $x_0(\nu^{-2} - 1)^{-1/2}$ | $\dfrac{(\Delta N_{max})(z_0/x_0)}{\pi r_0 A}$ |
| ⊥ prolate 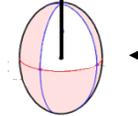 | 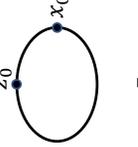 | $A = x_0/z_0$ | $z_0(\nu^{-2} - 1)^{-1/2}$ | $\dfrac{\Delta N_{max}}{\pi r_0}$ |
| ∥ oblate ($i = 0$ view) 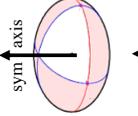 | 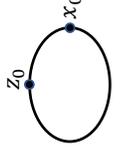 | $A = \dfrac{\cos i}{[(z_0/x_0)^2 - (\sin i)^2]^{1/2}}$ | $(x_0/A)(\nu^{-2} - 1)^{-1/2}$ | $\dfrac{(\Delta N_{max})(z_0/x_0)}{\pi r_0}$ |
| ⊥ oblate 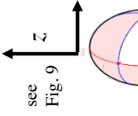 see Fig. 9 | 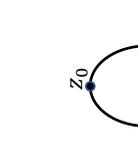 | $A = z_0/x_0$ | $x_0(\nu^{-2} - 1)^{-1/2}$ | $\dfrac{\Delta N_{max}}{\pi r_0 A}$ |

$A$, $r_0$ and $n_0$ are initial guesses for column density formulas in Table 2. The sym axis inclination $i$ from POS to observer must be assumed, e.g. $i = \pi/4$.

When the lowest closed column density contour extends many scale lengths from the peak, i.e. when $\omega_l \gg 1$, the term $1 - \omega_l^{-1} \tan^{-1} \omega_l$ in equation (4) can be approximated as 1, and the term $(1 + \omega_l^2)^{1/2}$ in equation (13) can be approximated as $\omega_l$. Then equation (13) can be rewritten in simpler form as

$$1 + 3\chi = 4\beta\chi^3 \qquad (14)$$

where $\chi$ is a scale factor which relates the density contrast ratio $v_0 = n_0/n_u$ and the normalized map radius $\omega_l$,

$$\chi \equiv v_0/\omega_l^2 , \qquad (15)$$

and where $\beta \equiv C/C_{\min}$, $C \equiv \epsilon_l/(v_{Nl} - 1)^3$, and $C_{\min} \equiv 4[4/(3\pi)]^3 = 0.306$.

The scale factor $\chi$ is found by solving the cubic equation (14). Of its three roots, the one real root is

$$\chi = \frac{f}{2\beta} + \frac{1}{2f} \qquad (16)$$

where $f \equiv \left\{\beta^2 + [(\beta - 1)\beta^3]^{1/2}\right\}^{1/3}$. This solution is physically realistic when $0 \leq \chi \leq 1$, which requires $1 \leq \beta \leq \infty$. Note that $\beta = 1$ implies $f = 1$ and $\chi = 1$. The approximation $\omega_l \gg 1$ produces an uncertainty in $\chi$ less than 5% provided $\omega_l > 5$, according to numerical calculations using equations (11) and (12).

## 2.7. Background Parameters

In equation (16) the formation efficiency $\epsilon_l$ of the mass within the lowest contour and the ratio $v_{Nl}$ of the lowest-contour column density to the background column density appear only in the combination $\beta = (\epsilon_l/C_{\min})/(v_{Nl} - 1)^3$. Thus for a given value of $0 \leq \epsilon_l \leq 1$, the allowed range of $v_{Nl}$ is $1 \leq v_{Nl} \leq 1 + [\epsilon_l/(C_{\min}\beta)]^{1/3}$. For a given value of $v_{Nl} \geq 1$, the allowed range of $\epsilon_l$ is $C_{\min}\beta(v_{Nl} - 1)^3 \leq \epsilon_l \leq 1$. When possible, $\epsilon_l$ and $v_{Nl}$ should be estimated for each observed map, to combine into $\beta$. Otherwise, the following points can be used to guide the allowed choices of $\epsilon_l$, $v_{Nl}$, and $\chi$. These choices follow the constraints that $\chi$ is a positive real solution to equation (16), that $\epsilon_l \leq 1$, and that $v_{Nl} \geq 1$.

### 2.7.1. Value of $\chi$

In the allowed range of $\chi$ from 0 to 1, values of $\chi$ become more probable as $\chi$ approaches 1, assuming that all allowed values of $v_{Nl}$ and $\epsilon_l$ are equally likely. This property follows because $\chi$ depends only on $\epsilon_l$ and $v_{Nl}$, and because the probability density $p(\chi)$ is proportional to the allowed area in the $v_{Nl} - \epsilon_l$ plane. The above definition of $\beta$ indicates that the allowed differential area between $\epsilon_l$ and $\epsilon_l + d\epsilon_l$ is $dA_{v\epsilon} = d\epsilon_l[\epsilon_l/(C_{min}\beta)]^{1/3}$. Integration over the allowed range of $\epsilon_l$, $0 \leq \epsilon_l \leq 1$, gives the allowed area, $A_{v\epsilon} = (3/4)(C_{min}\beta)^{-1/3}$. Then as $\beta$ decreases toward its minimum value of 1, $\chi$ increases toward its maximum value of 1, while $A_{v\epsilon}$ increases toward its maximum value $A_{v\epsilon,max} = (3/4)C_{min}^{-1/3} = 1.34$. Therefore the most probable value of $\chi$ is its maximum value, $\chi = 1$.

### 2.7.2. Values of $\epsilon_l$ and $v_{Nl}$

This section gives examples where the formation efficiency is $\epsilon \approx 0.3$ and where $v_{Nl} \approx 2$. Although these are plausible examples, it remains to be determined what values are most representative of appropriately selected star-forming regions. It is possible to estimate $v_{Nl}$ when it is a fit parameter in column density modelling, as shown in Section 5.1.

In the star-forming clump NGC 1333, the core mass fraction can be estimated as $\epsilon \approx 0.3$, since the clump has mass 568 $M_\odot$ for column density $N \geq 5\ 10^{21}$ cm$^{-2}$, and since it contains 24 starless cores with 112 $M_\odot$ and 18 protostellar cores with 71 $M_\odot$ (Sadavoy et al. 2010, Mercimek et al. 2017). The resulting mass fraction in cores is $\epsilon = 183/568 = 0.32$.

On smaller scales where filaments and cores can be approximated as idealized isothermal bodies, $\epsilon \approx 0.3$ approximates the mass ratio of a chain of critical cores having Jeans-length spacing to the mass of a critical parent filament. The mass ratio of a critical Bonnor-Ebert sphere to a Jeans-length segment of a critically stable filament is $\epsilon = 1.18/(2\pi^{1/2}) = 0.33$, when the sphere and filament have the same velocity dispersion and the same mean density (McKee & Ostriker 2007; hereafter MO07).

The column density ratio $v_{Nl} \approx 2$ was found in the filamentary clouds B211 and B213 observed by the *Herschel Space Observatory*. Their column density maps were fit to a model column density profile expected for a Plummer density function with index close to $p = 2$, added to a slowly varying background (Palmeirim et al. 2013). The background column density $N_u$ was defined for projected distance $r$ from the filament axis greater than 0.4 pc, where the slope of the column density profile $N(r)$ is noticeably shallower than for $r < 0.4$ pc. Then $N_u$ varies with $r$ either as a constant on the SW side, or as the sum of a constant and linearly increasing term on the NE side. At $r = 0.4$ pc the ratio of filament column density to the mean background column density is then $v_{Nl} = 2.2$ on the SW side and $v_{Nl} = 1.9$ on the NE side.

### 2.7.3. Choice of $\chi$

In the above examples the parameters of the background model are close to $\epsilon_l = 1/3$ and $v_{Nl} = 2$. These values are well-matched when $\beta = \chi = 1$, as favored by the probability discussion in Section 2.7.1. Then equation (15) relates the background density $n_u$ to the peak density $n_0$ and the dimensionless map radius $\omega_l$ by

$$n_u = n_0 \omega_l^{-2} \quad . \tag{17}$$

To illustrate how these choices give the background density and column density, consider a spheroidal condensation whose column density map gives peak density $n_0 = 10^5$ cm$^{-3}$, scale length $r_0 = 0.01$ pc, and lowest-contour short-axis radius $b_l = 0.10$ pc from the relations in Table 1. Then $\omega_l = b_l/r_0 = 10$, so equation (17) gives the background density $n_u = 10^3$ cm$^{-3}$ and the density contrast ratio $v_0 = n_0/n_u = 100$. Similarly, equation (12) gives the background column density once the aspect ratio is specified. For $A = 2$, equation (12) gives $N_u = 1.8 \times 10^{21}$ cm$^{-2}$.

### 2.8. Field Direction Models

The field angle $\theta_B(\xi, \zeta)$ in the plane of the sky ($\eta = 0$) was given in Paper 1 equation (26), for the sphere whose magnetic axis lies in the plane of the sky. This expression was obtained from the slope $d\zeta/d\xi$ of a flux tube wall in the plane of the sky. It is identical to that obtained from the magnetic field components given in M66 equations (11) and (12),

$$\theta_B(\xi,\zeta) = \tan^{-1}\left(\frac{1-t}{s^{-1}+st}\right) \quad . \tag{18}$$

Here $t$ is the ratio of density to mean density at $(\xi,\zeta)$ in equation (3), and $s \equiv \xi/\zeta$ is the ratio of the horizontal and vertical coordinates. The field angle $\theta_B$ is defined in spherical polar coordinates as in M66, as noted in Section 2.1. In the plane of the sky $\theta_B$ increases to the W of N, in contrast to the conventional position angle, which increases to the E of N.

This section derives simple modifications to equation (18) to give a more general expression for field angle in equation (23), for the two prolate and two oblate spheroids discussed above. Initial guesses for angle model fitting are discussed in Section 2.9.

### 2.8.1. Field Directions for Spheroids with Magnetic Axes in the Plane of the Sky

For a perpendicular prolate spheroid or parallel oblate spheroid, contours of constant density in the $x$-$z$ plane are ellipses elongated in the $x$-direction, with aspect ratio $A$. When the magnetic axis lies in the $z$-direction in the plane of the sky, the dimensionless spheroidal radius of a point at $(\xi,\zeta)$ has the same expression for $\omega = [(\xi/A)^2 + \zeta^2]^{1/2}$ in both the oblate and prolate case since $\eta = 0$. When the slope of a flux tube in the plane of the sky is computed to obtain the polar angle of the field direction $\theta_B$, the dependence of $\omega$ on $A$ modifies the spherical expression for $\theta_B$ in equation (10) to

$$\theta_B(\xi,\zeta) = \tan^{-1}\left(\frac{1-t}{s^{-1}+stA^{-2}}\right) \tag{19}$$

Equation (19) shows that as $A$ increases, $\theta_B$ at the same position increases, as is also evident from comparing Figures 1 and 3 for $\nu_0 = 30$ in Paper 1.

For a parallel prolate spheroid or a perpendicular oblate spheroid, contours of constant density in the $x$-$z$ plane are ellipses elongated in the $z$-direction, rather than in the $x$-direction as above. If the magnetic axis lies in the plane of the sky, $\omega = [\xi^2 + (\zeta/A)^2]^{1/2}$. The magnetic field angle is derived in the same way as for equation (19), but now it decreases with increasing $A$, according to

## Table 2
Column density $\Delta N(\xi,\zeta)$ and magnetic field direction $\theta_B(\xi,\zeta)$ in $p=2$ Plummer spheroids inclined through angle $i$

| Spheroid type | aspect ratio | $\Delta N(\xi,\zeta)$ | $q$ in eq. (23) | $\omega^2$ in eq. (5) for $t_q$ in eq. (23) |
|---|---|---|---|---|
| ∥ prolate ($i=0$ view) | $A$ | $\dfrac{\pi n_0 r_0 A}{\{(1+\xi^2)[1+(A^2-1)(\cos i)^2]+\zeta^2\}^{1/2}}$ | $A^{-1}\sec i$ | $\xi^2 + [(\zeta/A)\sec i]^2$ |
| ⊥ prolate | $A$ | $\dfrac{\pi n_0 r_0}{[1+(\xi/A)^2+\zeta^2]^{1/2}}$ | $A\sec i$ | $(\xi/A)^2 + (\zeta \sec i)^2$ |
| ∥ oblate ($i=0$) | $A$ | $\dfrac{\pi n_0 r_0 A}{\{(1+(\xi/A)^2)[1+(A^2-1)(\sin i)^2]+\zeta^2\}^{1/2}}$ | $A\sec i$ | $(\xi/A)^2 + (\zeta \sec i)^2$ |
| ∥ oblate, twist | $A$ | same as for ∥ oblate | $A\sec i \coth(\zeta/\alpha)$ | $\{(\xi/A)\coth[(\zeta/\alpha)\sec i]\}^2 + (\zeta \sec i)^2$ |
| ⊥ oblate | $A$ | $\dfrac{\pi n_0 r_0 A}{[1+\xi^2+(\zeta/A)^2]^{1/2}}$ | $A^{-1}\sec i$ | $\xi^2 + [(A)\sec i]^2$ |

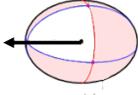
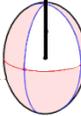
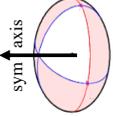
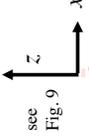
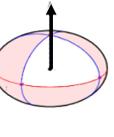

Normalized coords $\xi \equiv x/r_0$; $\zeta \equiv z/r_0$; $s \equiv x/z$; $\nu_0$ = peak/background density $n_0/n_u$; $\theta_B$ and $i$ are polar angles from $z$-axis; $\alpha$ = twist scale height

$$\theta_B(\xi,\zeta) = \tan^{-1}\left(\frac{1-t}{s^{-1}+stA^2}\right) \quad . \tag{20}$$

In each of these cases the dependence of field angle on aspect ratio can be understood as arising from stretching the field pattern for a centrally condensed sphere. When the sphere is stretched perpendicular to its magnetic axis, the field angle at a given position increases; when it is stretched parallel to the magnetic axis, the angle decreases. This behavior is shown in Paper I, Figures 1 and 4 for $v_0 = 30$.

### 2.8.2. Field Directions for Spheroids with Inclined Magnetic Axes

Equations (19) and (20) give the field angle in a plane which contains the magnetic axis of the spheroid. For convenience this plane is assumed to coincide with the plane of the sky. If the spheroid and its associated symmetry plane are inclined through angle $i$ toward the observer, equations (19) and (20) can be modified to give the projection of each field line in the inclined plane back onto the plane of the sky. The projected height of a point on an inclined plane is foreshortened compared to its uninclined height, while its horizontal position is unchanged by the inclination. Thus a point on the inclined plane with coordinates $(\xi,\zeta)$ has uninclined coordinates $(\xi_0,\zeta_0)$ related by $(\xi,\zeta) = (\xi_0, \zeta_0 \cos i)$. Similarly, the tangent of the field polar angle at the inclined point and at the uninclined point are related by $\tan \theta_B(\xi,\zeta) = (\sec i) \tan \theta_B(\xi_0,\zeta_0)$. Then for a perpendicular prolate or parallel oblate spheroid of aspect ratio $A$, applying equation (19) for the field angle at an uninclined point gives the field angle at the inclined point as

$$\theta_B(\xi,\zeta) = \tan^{-1}\left[\frac{1-t_0}{s^{-1}+st_0(A \sec i)^{-2}}\right] \tag{21}$$

where $t_0$ is the density ratio $v/\bar{v}$ evaluated at the uninclined point $(\xi_0,\zeta_0) = (\xi, \zeta \sec i)$. For a parallel prolate or parallel oblate spheroid, $A$ in equation (21) is replaced by $1/A$.

This estimate treats the projection of field lines in an inclined plane onto the plane of the sky, but does not take into account the differing projections of the front and rear of a flux tube. A more detailed discussion of projection of an inclined flux tube is given in Kataoka et al (2012) and in Paper I.

### 2.8.3. Field Directions for Oblate Spheroids with a Toroidal Twist

In star-forming disks, the magnetic field pattern may acquire a toroidal component due to rotational stretching and twisting of field lines, in addition to the poloidal component which arises from flux freezing at larger scales (Tomisaka 2011). The toroidal component is related to the efficiency of magnetic braking, and to the magnetic driving of outflows (Joos et al. 2012, Tomida et al. 2015, Masson et al. 2016). Toroidal fields are expected to be more significant during the protostellar accretion phase than during the preceding infall phase. They may extend for up to ~ 200 au in radius in simulations of low-mass star formation (Masson et al. 2016). They may extend farther, according to ALMA polarization observations of HH 211. These observations have been interpreted to show a toroidal component in a rotating pseudodisk of radius ~ 400 au (Lee et al. 2019). At still larger scales, the toroidal component decreases, and the field pattern transitions to the more poloidal hourglass shape.

A simple analytic representation is used here, as in Padovani et al. (2013; hereafter P13) rather than an exact physical model. It is assumed that an oblate spheroid has symmetry axis in the $z$-direction, and that its initial poloidal field lines have coordinates $(\xi_p, \zeta_p)$ in the plane of the sky. Rotation about the $z$-axis is assumed to stretch and twist the initial field lines in the plane of the sky into a more toroidal configuration. Their azimuthal excursion is assumed to be greatest at the equator and to fall off with increasing vertical distance. The scale height $a$ of the fall-off is expected to be comparable to the disk scale height, and is denoted $\alpha \equiv a/r_0$ in dimensionless form. The maximum azimuthal excursion is assumed to be $\pi/2$ radians, so that a point on an initially poloidal field line at $(\xi_p, 0)$ travels to a projected position $(\xi, \zeta) = (0,0)$. Field lines which similarly approach the origin approximate the configuration for the accretion phase of a rotating, collapsing, magnetized sphere, as shown in Figure 5 of Tomisaka (2011).

If the vertical fall-off of the poloidal field component is exponential in form, the projected toroidal and poloidal coordinates can be written in terms of the initial poloidal coordinates as $[\xi, \zeta] = [\xi_p \tanh(\zeta_p/\alpha), \zeta_p]$. Equivalently, the untwisted position can be written $(\xi_p, \zeta_p) = [\xi \coth(\zeta/\alpha), \zeta]$. Initial poloidal field lines and twisted toroidal and poloidal field lines calculated with this procedure have distinctly different shape near the origin. The poloidal lines are vertical near the origin, as in P13 Figure 2 (top), while the combined toroidal and poloidal converge radially toward the origin, as in P13 Figure 2 (middle). At large distance from the origin the two patterns become identically poloidal. This approximate formulation is intended for assigning and comparing fit parameters to observed polarization patterns.

The field angle $\theta_B$ in a toroidally twisted field pattern can now be derived in a similar way to that for the inclined field axis in equation (21). The tangent of the field angle at a twisted point is related to the tangent of the field angle at the corresponding untwisted poloidal point by $\tan \theta_B(\xi,\zeta) = \tanh(\zeta_p/\alpha)\tan\theta_B(\xi_p,\zeta_p)$. Then if a spheroid of aspect ratio $A$ has field axis inclined by $i$ from the vertical, and its untwisted field angle is described by equation (21), the field angle at the twisted position can be written as

$$\theta_B(\xi,\zeta) = \tan^{-1}\left\{\frac{1-t_p}{s^{-1}+st_p[A\sec i\coth(\zeta/\alpha)]^{-2}}\right\} \qquad (22)$$

where $t_p$ is the density ratio $v/\bar{v}$ evaluated at the untwisted, uninclined poloidal position $(\xi_p,\zeta_p) = [\xi\coth(\zeta/\alpha), \zeta\sec i]$.

The foregoing analysis may also be applied to a differentially rotating parallel prolate condensation using the substitution of $A^{-1}$ for $A$ as discussed in Section 2.8.1.

### 2.8.4. Field Directions for Elongated, Inclined Spheroids with a Toroidal Twist

Summarizing the polar angle expressions in equations (18) - (22), the field angle can be written

$$\theta_B(\xi,\zeta) = \tan^{-1}\left(\frac{1-t_q}{s^{-1}+st_q q^{-2}}\right) \qquad (23)$$

where $q$ is the product of the factors representing spheroid elongation [$A$ or $A^{-1}$], magnetic axis inclination [$\sec i$], and rotational twisting of poloidal field lines [$\coth(\zeta/\alpha)$], and $t_q$ is the corresponding density ratio calculated from equation (4). The expressions for $q, t_q$, and for their arguments are discussed below, and they are summarized in Table 2.

Then for elongation alone ($A > 1, i = \alpha = 0$), $q = A$ or $A^{-1}$ and $t_q = t(\xi,\zeta)$. For elongation and inclination ($A > 1, i > 0, \alpha = 0$), $q = A\sec i$ or $A^{-1}\sec i$ and $t_q = t(\xi, \zeta\sec i)$. For elongation and twisting ($A > 1, i = 0, \alpha > 0$), $q = A\coth(\zeta/\alpha)$ or $A^{-1}\coth(\zeta/\alpha)$, and $t_q = $ $t(\xi \coth(\zeta/\alpha), \zeta)$. For elongation, inclination, and twisting ($A > 1$, $i > 0$, $\alpha > 0$), $q = A \coth(\zeta/\alpha) \sec i$ or $A^{-1} \coth(\zeta/\alpha) \sec i$ and $t_q = t(\xi \coth(\zeta/\alpha), \zeta \sec i)$. Increasing elongation perpendicular to the field, or decreasing elongation parallel to the field, or increasing $i$ or $\alpha$ tends to increase the magnetic field angle $\theta_B$.

Table 2 summarizes equation (23) and gives expressions for $q(\omega)$, $t_q(\omega)$, and $\omega^2$ for each of the four spheroid types considered. In Table 2, the expressions for $\omega^2$ in $\theta_B(\omega)$ differ from the standard spheroid expressions for $\omega^2$ in $\nu(\omega)$ and $\bar{\nu}(\omega)$ in Section 2, because $\omega^2$ in $\theta_B(\omega)$ includes the factors which relate the projected inclined and twisted field directions to their uninclined, untwisted directions. The expressions for $\omega^2$ in $\theta_B(\omega)$ reduce to those for $\omega^2$ in $\nu(\omega)$ and $\bar{\nu}(\omega)$ when $i = \alpha = 0$. The expressions in Table 2 also show that $\theta_B(x,z)$ in equation (23) for a Plummer spheroid of density contrast $\nu_0$ and any elongation, inclination, and twist reduces to equation (18) for a Plummer sphere of the same $\nu_0$, whose field is uninclined and untwisted, when $A = 1$, $i = 0$, and $\alpha = 0$.

To illustrate results of Section 2.3 - 2.7, Figure 1 shows the progression of spheroid column density and field direction models, as an uninclined Plummer sphere with purely poloidal field lines is altered by elongation, inclination, and toroidal twisting. There, maps are based on a $p = 2$ Plummer model star-forming envelope having peak density $n_0 = 3 \times 10^{10}$ cm$^{-3}$ and scale length $r_0 = 30$ au, following the density range from 1 au to 1000 au in the simulation of a rotating, magnetized protostellar envelope at the epoch of second core formation (Vaytet et al. 2018).

In Figure 1, the upper left panel (*a*) shows the initial spherical case with magnetic axis in the plane of the sky. The flux tube lines have spacing following Section 2.4.3 of Paper 1. The upper right panel (*b*) shows an oblate spheroid with aspect ratio $A = 2$ and with all other parameters as in (*a*). The lower left panel (*c*) shows the spheroid in (*b*) after inclination through 45 degrees, and the lower right panel (*d*) shows the spheroid in (*c*) with a toroidal twist for a scale height of 150 au ($\alpha = 5$). The column density morphologies in (*a*), (*b*), and (*c*) are distinctly different due to their changes in elongation and inclination; (*c*) and (*d*) are equal by assumption. The field line morphologies (*a*) - (*d*) follow a sequence of increasing field angles from the vertical direction, as expected from the discussion in Section 2.7.

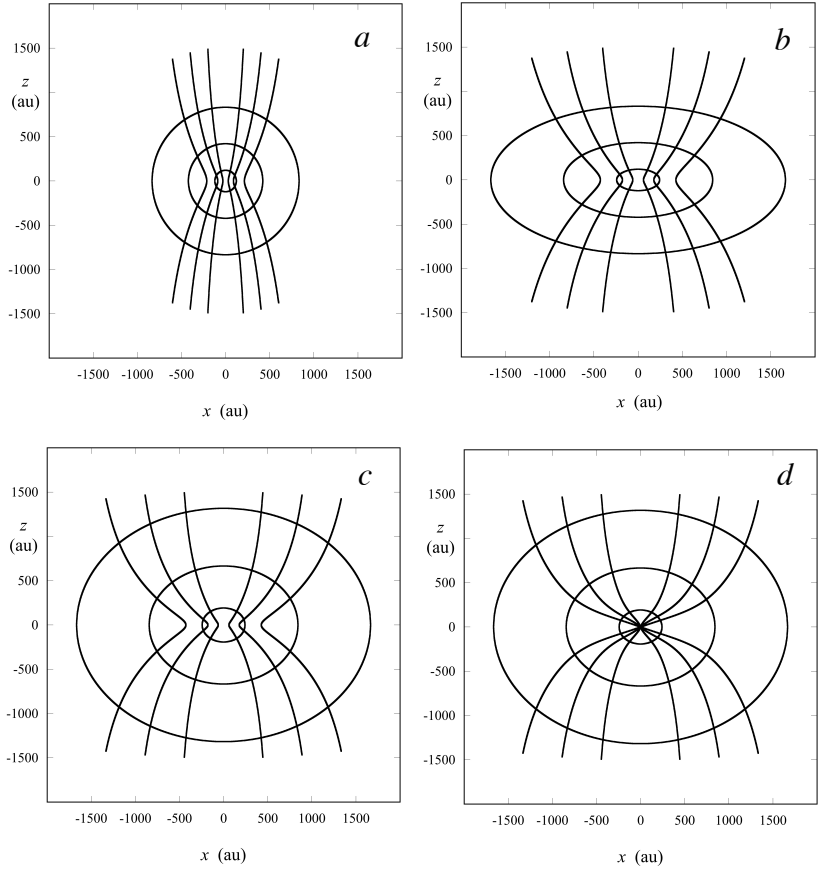

**Figure 1.** Contours of constant column density and flux tube lines for a circumstellar envelope modelled as a $p = 2$ Plummer oblate spheroid, with increasing aspect ratio, inclination, and toroidal twist. The spheroid has scale length $r_0 = 30$ au, peak density $3 \times 10^{10}$ cm$^{-3}$ and background $3 \times 10^7$ cm$^{-3}$. The four panels (*a*) - (*d*) represent values of aspect ratio $A$, inclination angle $i$, and normalized twist scale height $\alpha$, progressing from $(A, i, \alpha) = (1, 0, 0)$ to $(2, 0, 0)$ to $(2, \pi/4, 0)$ to $(2, \pi/4, 5)$. The column density contours are drawn at 1.5, 3.0, and $10 \times 10^{24}$ cm$^{-3}$.

## 2.9. Field Direction Parameters

The parameters $A$, $r_0$, and $v_0$ obtained from Sections 2.5-2.7 can be used as inputs to the $\theta_B$ model, equation (22), to fit an observed polarization map. The remaining input parameters are the magnetic axis inclination $i$, and the dimensionless scale height of the toroidal twist, $\alpha$. Alternately some of the parameters $A$, $r_0$, and $v_0$ may be considered as free parameters for fitting to the polarization map, independent of the column density model.

An initial guess for $i$ may be obtained if the column density fit depends on $i$, as for the parallel prolate and parallel oblate spheroids. The density structure and field line structure can be expected to incline together since the field lines are frozen into the gas. For the perpendicular prolate and perpendicular oblate spheroids, no preferred initial guess for $i$ is available, and the mid-range value $\pi/4$ may be used. For $\alpha$, the normalized value of a few disk scale heights may be used.

## 3. MAGNETIC FIELD STRENGTH MODEL

This section gives expressions for the mass-to-flux ratio and for the field strength as a function of position, combining an estimate of the mean field strength with the foregoing density model.

### 3.1. Mean Field Strength

This section estimates the mean field strength within the radius of a condensation where the nonthermal component of the line-of-sight velocity dispersion $\sigma_{NT}$ is related to Alfvénic fluctuations, following DCF and Ostriker et al. (2001, hereafter OSG01); and Auddy et al. (2019, hereafter A19).

In a cloud of uniform density $\rho$, the observed nonthermal velocity dispersion $\sigma_{NT}$ along the line of sight consists of motions perpendicular to the plane-of-the-sky component of the field $B_p$. Then the field strength associated with Alfvén waves can be expressed in terms of $\rho$ and the relative wave amplitude $\sigma_{B_p}/B_p$ by

$$B_p = \frac{Q_{B_p}(4\pi\rho)^{1/2}\sigma_{NT}}{\sigma_{B_p}/B_p} \quad . \tag{24}$$

The term $\sigma_{B_p}/B_p$ can be equated to the dispersion $\sigma_\theta$ of plane-of-sky polarization angles, assuming that the mean polarization direction for grain emission is perpendicular to the mean direction of $B_p$ and that the efficiency of grain alignment is perfect. Then equation (24) is the basis of the DCF method of field estimation (Davis (1951), Chandrasekhar & Fermi (1953), OSG01, A19). Here the coefficient $Q_{B_p}$ has the same meaning as $Q$ in OSG01 equation (16). The notation $Q_{B_p}$ distinguishes this coefficient for the plane-of-the-sky field strength from the coefficient $Q_B$ for the total field strength, discussed below.

In equation (24), the analytic estimate of Chandrasekhar & Fermi (1953) has $Q_{B_p} = 1$, but a simulation of magnetized, turbulent molecular clouds matches equation (24) only if $Q_{B_p} \approx 1/2$ due to more complex field and density structure. This match also applies only for the range of inclination angles $0 \leq i \lesssim 60$ deg between the mean field direction and the plane of the sky, and only for field fluctuation amplitudes in the linear regime $\sigma_{B_p}/B_p \lesssim 0.5$ (OSG01; see also Kudoh & Basu 2003 and A19). Equation (24) with $Q_{B_p} = 1/2$ is widely used to estimate $B_p$.

It is useful to estimate the total field strength $B = B_p \sec i$ corresponding to $B_p$, to better estimate the mass-to-flux ratio and field strength structure of a condensation. In some protostellar regions, $B$ may be estimated directly from $B_p$ if the mean field direction can be assumed to align with the known direction of the associated outflow axis. Otherwise, the following statistical estimate can be made. Assuming that the inclination of the field direction is equally likely over the range of $i$ where equation (24) applies, the mean field strength over this range is $\langle B \rangle = I B_p$ where $I \equiv \int_{i_{\min}}^{i_{\max}} di \, \sec i / (i_{\max} - i_{\min}) = 1.26$, with corresponding standard deviation $\sigma_I = 0.25$. This factor of 1.26 has essentially the same value as $4/\pi$, the factor given by Heiles & Crutcher (2005) and by Pillai et al. (2016).

Applying equation (24) to the spatial average field strength $\bar{B}_1$ within a particular dimensionless radius $\omega_1$ of a condensation map, and substituting $\langle B \rangle = B_p I$ gives

$$\langle \bar{B}_1 \rangle = Q_B \overline{(B/\sigma_B)}_1 (4\pi \bar{\rho}_1)^{1/2} (\bar{\sigma}_{NT})_1 \qquad . \qquad (25)$$

Here the notation $\langle x \rangle$ indicates the inclination-average of the dummy variable $x$, while $\bar{x}_1$ and $(\bar{x})_1$ each indicate the spatial average of $x$ within radius $\omega_1$. The coefficient $Q_B \pm$ its associated uncertainty $\sigma_{Q_B}$ can be written $Q_B \pm \sigma_{Q_B} = (I \pm \sigma_I) Q_{B_p} = 0.63 \pm 0.13$. It is assumed that the relative wave amplitude is independent of inclination, i.e. $\sigma_B/B = \sigma_{B_p}/B_p$, and that $\overline{\rho^{1/2}} \approx (\bar{\rho})^{1/2}$. This latter approximation is useful because $(\bar{\rho})^{1/2}$ can be expressed analytically via equation (4). The approximation is sufficiently accurate since $\overline{\rho^{1/2}}$ and $(\bar{\rho})^{1/2}$ have a mean deviation of $< 3\%$ for $0 \leq \omega \leq 30$ and $\nu_0 = 100$.

In practical units equation (25) can be written

$$\langle \bar{B}_1 \rangle = [88 \pm 18] \left[\frac{\overline{(B/\sigma_B)_1}}{2}\right] \left[\frac{\bar{n}_1}{10^4 \text{ cm}^{-3}}\right]^{1/2} \left[\frac{(\bar{\sigma}_{NT})_1}{\text{km s}^{-1}}\right] \mu G \qquad (26)$$

where $\bar{n}_1$ is the mean volume density within $\omega_1$. This estimate can be further refined if the dispersion of random polarization angles within $\omega_1$ is known. Henceforth this dispersion in observed angles is denoted $\sigma_{\Delta\theta}$ to account for subtraction of a model of the organized component, to reveal the random component. Then $\overline{(B/\sigma_B)}_1 = (1/\sigma_{\Delta\theta})_1$ can be substituted in equations (25) and (26), as discussed in Section 4.5.

### 3.2. Mass-to-Critical-Mass Ratio

The ratio of mass to magnetic critical mass indicates whether the self-gravitating mass of a condensation is supercritical, i.e. it is high enough to overcome magnetic forces and allow collapse $(M/M_c > 1)$; or whether its mass is subcritical, or too low enough to allow collapse $(M/M_c < 1;$ MO07). For a sphere of radius $r_1$ and mean mass density $\bar{\rho}_1$ this ratio can be written

$$\frac{M}{M_c} = \frac{4\bar{\rho}_1 r_1 G^{1/2}}{3 c_\Phi \langle \bar{B}_1 \rangle} \qquad (27)$$

where the coefficient $c_\Phi = 1/(2\pi)$ is the standard value for an infinite cold sheet (Nakano & Nakamura 1978), and where the coefficient for other geometries differs by a factor less than two (Strittmatter 1966, Nakano & Nakamura 1978).

The DCF estimate of mean field strength in a condensation of given mass and flux does not depend on the SFF assumption that the mass formed from an initial uniform medium while conserving mass and flux. However, since $M/M_c$ is assumed constant in time for a SFF condensation, the constraints on its past contraction depend on $M/M_c$ in the same way as on its future contraction. If $M/M_c > 1$ the condensation could have formed by self-gravity overcoming magnetic forces, so the DCF and SFF models are consistent. If $M/M_c < 1$ magnetic forces would have prevented gravitational collapse, and the DCF and SFF models are inconsistent. In this case, the spheroid formation would have required an increase in mass due to flow along field lines, or a reduction in flux due possibly to reconnection or ambipolar diffusion (MO07). In BHR71 IRS1, $M/M_c > 1$ according to Section 5. In that case the DCF and SFF models are consistent.

### 3.3. Field Strength Structure

The structure of the field strength in the equatorial plane perpendicular to the magnetic axis can be written in terms of the mean field strength within a particular dimensionless radius given in equation (25), following M66 and Paper 1,

$$B(\omega) = \langle \bar{B}_1 \rangle \frac{\nu(\omega)}{[\bar{\nu}(\omega)]^{1/3}[\bar{\nu}(\omega_1)]^{2/3}} \quad . \tag{28}$$

This relation assumes that the mass within $\omega_1$ is supercritical, i.e. it is greater than the magnetic critical mass within $\omega_1$, as discussed in Section 3.2 above. When the magnetic axis is inclined through polar angle $i \geq 0$ from the $z$-axis toward the observer in the $y$-$z$ plane, equation (28) can be evaluated conveniently along the $x$- axis, since projected lengths in the $x$ -direction are independent of $i$. Then for the sphere and each of the four spheroids considered here, $\omega$ is evaluated with $\zeta = \eta = 0$, giving either $\omega = \xi$ for each prolate spheroid, or $\omega = \xi/A$ for each oblate spheroid.

Substitution of the appropriate expression for $\omega$ in equation (28) gives the continuous structure of field strength along the $x$-axis. It gives three expressions of particular interest. The uniform background field strength $B_u$ when $\omega \gg 1$ is

$$B_u = \langle \bar{B}_1 \rangle [\bar{\nu}(\omega_1)]^{-2/3} \tag{29}$$

using $\nu(\omega \gg 1) = \bar{\nu}(\omega \gg 1) = 1$. The field strength $B$ in terms of $B_u$ is

$$B(\omega) = B_u \nu(\omega)[\bar{\nu}(\omega)]^{-1/3} \quad , \tag{30}$$

and the peak field strength $B_0$ when $\omega \ll 1$ is

$$B_0 = B_u(1 + \nu_0)^{2/3}, \tag{31}$$

using $\nu(\omega \ll 1) = \bar{\nu}(\omega \ll 1) = 1 + \nu_0$. For all positions in the equatorial plane, including positions along the x- axis, the magnetic field direction is perpendicular to the equator, with unit vector $\hat{B} = \hat{z}\cos i - \hat{y}\sin i$.

## 4. STEP-BY-STEP APPLICATION

This section summarizes the steps needed to obtain magnetic properties from the observed maps and from the models of column density and field morphology, described in Sections 2 and 3.

### 4.1. Setup

For each pair of column density and polarization maps, the spheroid type (parallel prolate, perpendicular prolate, parallel oblate, or perpendicular oblate) is chosen by inspection of the maps and by other information (e.g. flattened circumstellar envelopes are oblate rather than prolate, filaments with aspect ratio > ~3 are prolate rather than oblate). The polarization map symmetry axis and the projected magnetic axis are assigned to the z - axis. It may be convenient to rotate the maps on the plane of the sky, as in Section 2.2, so that the z - axis is either vertical, or it is aligned with the projected outflow axis.

### 4.2. Column Density Model

The column density map is used to obtain the corresponding column density model $N(x, z)$ for a $p = 2$ Plummer spheroid in a uniform medium, as in Paper 1. Estimates of the peak density $n_0$, scale length $r_0$, and aspect ratio $A$ are obtained from the column density map contours, using the expressions in Table 1. The background column density $N_u$ and background density $n_u$ are estimated from the radius and column density of the lowest closed contour of column density and an assumed value of the parameter $\beta$, as in Sections 2.6-2.7. Limitations due to uncertainty in the lowest contour are discussed in Section 6.3. The column density model is specified by using the above parameters and the inclination $i$ with the appropriate column density equation in Table 2 or in equation (7), (8), (9), or (10). When the model and map are sufficiently similar, the parameter values may be optimized by least-squares fitting of the model to the map.

## 4.3. Density Model

The normalized model density $v$, mean density $\bar{v}$, and density ratio $t = v/\bar{v}$ are obtained from equations (3) - (5) with the parameters of density $n_0$, scale length $r_0$, aspect ratio $A$, and background density $n_u$ obtained from the column density model.

## 4.4. Field Direction Model

The field direction model $\theta_B(x,z)$ is obtained from equation (23) and from $v_0$, $r_0$, $A$, $i$, and $\alpha$, using the appropriate expressions for $t_q$ in equation (5) and for $q$ in Section 2.6 and Table 2. The parameters $v_0$, $r_0$, and $A$ have values set in the column density and density models. If they are needed, the inclination $i$ and toroidal scale height $\alpha$ may be assumed, estimated from other observations, or estimated from fitting with the field direction model. In the ideal case, parameters which appear in both the column density model and the field direction model can be obtained consistently from fits to both models. In a more typical case, fitting to obtain well-determined parameter values may be successful only within particular zones of a map, and/or only when some parameters are free while others are held fixed.

## 4.5. Mean Field Strength

The mean field strength averaged over the condensation is based on equation (26). The density model parameters and equation (5) are used to obtain the mean density. The mean nonthermal velocity dispersion is obtained from observations of a well-resolved, optically thin spectral line whose map extent is similar to that of the column density map, after removal of velocity gradients and the contribution from thermal motions. If no suitable polarization observations are available, the mean ratio of field strength to field fluctuation amplitude $\overline{B/\sigma_B}$ is assumed, with a nominal value $\overline{B/\sigma_B} \lesssim 2$.

If suitable polarization observations are available, the distribution of angle differences $\Delta\theta$ between observation and model may be analyzed as in the DCF method. If the standard deviation $\sigma_{\Delta\theta}$ is less than the maximum $\sigma_{\Delta\theta,max} \approx 29$ deg for linear fluctuations (OSG01), it is more realistic to use the value $\overline{B/\sigma_B} = 1/\sigma_{\Delta\theta}$, with $\sigma_{\Delta\theta}$ in radians, rather than the fiducial value $\overline{B/\sigma_B} = 2$ in equation (26).

**4.6. Mass-to-Critical-Mass Ratio and Field Strength Structure**

The ratio of mass to magnetically critical mass is obtained by evaluation of equation (27), using the above mean field strength with the density model parameters and equations (3) and (4) for the density and mean density. When the mean field strength is based on the dispersion between model and observed polarization directions, it is essentially the same as the DCF field strength. The magnetic field structure in equations (28)-(31) is based on the same mean field strength and the same density model parameters as for the ratio of mass to magnetically critical mass.

**SECTION 5. MAGNETIC STRUCTURE OF THE ENVELOPE OF BHR71 IRS1**

This section applies the steps in Section 4 to derive magnetic field properties from observations of column density and polarization structure. The observed region is the envelope surrounding the protostar BHR71 IRS1, the brighter member of a wide binary pair in the globule BHR71 (Bourke et al. 1995, Bourke 2001). This region has been observed by ALMA at 1.3 mm wavelength in the dust continuum and in multiple lines, including $J = 2$-$1$ $C^{18}O$ (Tobin et al. 2019, hereafter T19), and in the unpolarized and polarized dust continuum (Hull et al. 2020, hereafter H20). These observations and the foregoing models give enough information to estimate the structure of the envelope column density, volume density, field strength and field direction. The observations probe spatial scales limited by the FWHM beam width of $1.5 \times 1.7$ arcsec, giving a linear resolution of ~300 au for estimated source distance 200 pc (T19; Bourke et al. 1997, Zucker et al. 2020).

**5.1. Column Density Structure**

The envelope column density structure is estimated from the ALMA map of integrated line intensity $W(C^{18}O) \equiv \int S_\nu(v) dv$ in the $J = 2$-$1$ line of $C^{18}O$ (T19 Figures 6 and 18). The $W(C^{18}O)$ map contours are centered on IRS1, they are centrally concentrated, and they are approximately circular in shape within ~1000 au. They are cospatial with the dust emission out to ~2000 au (T19 Figure 4). The 1.3 mm polarization also follows the $W(C^{18}O)$ map (H20 Figure 3), possibly because both gas and dust are heated by protostellar radiation, keeping the polarized signal bright and keeping $C^{18}O$ molecules in the gas phase (H20).

In many studies, the first choice for deriving column density structure is analysis of a suitable dust continuum intensity map. But in BHR71 IRS1 the point-like component of the continuum map is ~10 times brighter than the underlying envelope. This bright component may be point-like because with resolution ~300 au the disk emission is poorly resolved (T19). Deriving envelope structure in the presence of such a strong point source is possible (Frau et al. 2011), but it introduces additional uncertainty (Pokhrel et al. 2018). In contrast, the $W(C^{18}O)$ map has a well-defined local maximum but no point-like component. This property may arise if the unresolved disk gas has greater freeze-out than the resolved envelope gas. Significant localized freeze-out of CO in young disks is estimated to occur within "snow surfaces" extending from ~50 to ~200 au in radius and up to ~30 au in height (Qi et al. 2019 Figure 3). It is therefore assumed that the $W(C^{18}O)$ map traces the column density structure over the resolved scales 300 au - 5000 au, with less error and more convenience than does the continuum map.

The $C^{18}O$ gas in a protostellar envelope also has some abundance variation due to the structure of its freeze-out, and its emission in the $J = 2 - 1$ line is expected to be optically thick at its peak, and optically thin at large radius. These properties can nonetheless be approximated by assuming that the envelope of BHR71 IRS1 has constant $C^{18}O$ abundance over its resolved scales, and that its emission in the $J = 2 - 1$ line is optically thin. These assumptions greatly simplify the analysis of the observations. Despite their idealization, they provide an estimate of mean $C^{18}O$ abundance $[C^{18}O] \equiv N(C^{18}O)/N$, which has good agreement with the typical abundance for low-mass protostellar envelopes.

The $C^{18}O$ column density is related to the integrated line intensity $W(C^{18}O)$, using standard expressions for the partition function and column density (Goldsmith et al. 1997). For optically thin emission and rotation temperature 10 - 20 K, $N(C^{18}O) = 5.3 - 6.1 \times 10^{14}$ cm$^{-2}$ $W(C^{18}O)[\text{K km s}^{-1}]^{-1}$. The total gas column density and integrated intensity in the envelope of BHR71 IRS 1 are similarly related, by $N = 3.5 \times 10^{22}$ cm$^{-2}$ $W(C^{18}O)[\text{K km s}^{-1}]^{-1}$, from fitting the Plummer density model to the envelope map, in Figure 2 below. Then the mean $C^{18}O$ abundance is the ratio of these two expressions, or $[C^{18}O] = 2 \times 10^{-8}$. This abundance value lies near the peak of the $[C^{18}O]$ distribution in 15 Class 0 protostellar envelopes, according to observations of multiple lines and detailed envelope models (Yildiz et al. 2013, Table 5).

The assumptions of constant abundance and optically thin emission also imply that $N$ and $W(C^{18}O)$ are linearly proportional. This property is corroborated by the high-confidence fit of $N$ to $W(C^{18}O)$ in Figure 2, for the adopted Plummer model of $N$. There, the proportionality constant between $N$ and $W(C^{18}O)$ is determined with a relative uncertainty of 1%. Thus the assumptions of constant abundance and optically thin emission appear acceptable for analysis of the $C^{18}O$ envelope observations.

The BHR71 core harbors the binary system IRS1 and IRS2. The core mass $M = 4.6\ M_\odot$ and radius $R = 0.05$ pc (T19) are used to estimate the mass and mean column density of the IRS1 envelope. It is assumed that the BHR71 core and the IRS1 envelope each follow a $p = 2$ Plummer density profile in a local background as in Section 2.6, with $\omega_l \gg 1$. Then within the lowest IRS1 contour of radius $r_l = 4500$ au, the envelope mass is $M_l \approx (r_l/R)M = 2.0\ M_\odot$ and the mean column density is $\bar{N}_l = 9M_l/(16mr_l^2) = 1.3 \times 10^{23}$ cm$^{-2}$, including foreground and background gas.

The column density model parameters are derived by fitting the $p = 2$ Plummer column density profile to the observed $W(C^{18}O)$ profile. The $W(C^{18}O)$ profile is obtained from the contours of the $W(C^{18}O)$ map in T19 Figure 6 along a N-S line through the peak of the map. The model column density $N$ is the sum of the background column density $N_u = \pi n_0 r_0 (v_{Nl} - 1)^{-1}(1 + \omega_l^2)^{-1/2}$ and the background-subtracted column density $\Delta N$, using equation (8) with $A = 1$, in terms of quantities defined in Sections 2.3 and 2.6. The column density is normalized by the peak column density above background, $\tilde{N} \equiv N/(\pi n_0 r_0)$, and the integrated intensity is normalized by the rms noise, $\tilde{W} \equiv W(C^{18}O)/\sigma$ as in T19. Then $\tilde{W} = C_N \tilde{N}$, and the model fit function is

$$\tilde{W}_{\text{fit}} = C_N\left[(v_{Nl} - 1)^{-1}(1 + \omega_l^2)^{-1/2} + (1 + \omega^2)^{-1/2}\right]. \qquad (32)$$

Equation (32) has three free parameters determined by the fit: the scale factor $C_N$, the column density ratio $v_{Nl}$, and the scale length which normalizes $\omega$ and $\omega_l$. The radial profile data used for the fit are not corrected for beam smoothing, so each of the best-fit parameters is slightly different from its intrinsic value, which can be recovered by correction for smoothing. Of these parameters a significant correction will be needed only for the scale length, which is here denoted $r_{0s}$.

Figure 2 shows the $\tilde{W}$ data from T19 Figure 6 as a function of projected radius $r$, along with the best-fit model $\tilde{W}_{\text{fit}}$ from equation (32), and with the profile of the ALMA beam. The beam is approximated as a circularly symmetric Gaussian with 1-sigma radius 140 au, corresponding to the FWHM dimensions $1.7 \times 1.5$ arcsec (T19 table 1).

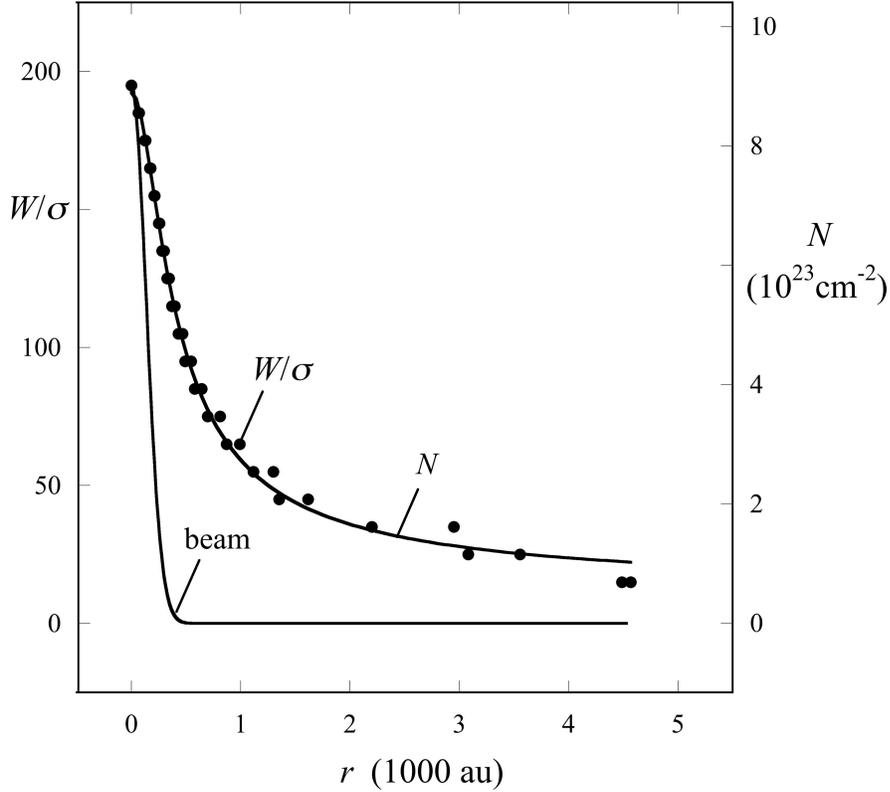

**Figure 2.** ALMA integrated intensity profile $W(C^{18}O)$ normalized by rms noise $\sigma$ (*filled circles, left scale*, T19) and best-fit model column density profile $N$ (*right solid line, right scale*), as functions of projected radius $r$. The beam profile (*left solid line*) is scaled to have the same maximum as $W(C^{18}O)$. The fit gives estimates and uncertainties of the column density scale factor $C_N$, the ratio $v_{Nl}$ of lowest-contour to background column density, and the density scale length $r_0$. The correlation coefficient of the fit is $R > 0.99$.

In Figure 2 the model fits the data well, with correlation coefficient $R > 0.99$, and with best-fit values ± one-sigma fit uncertainties $C_N = 181 \pm 2$, $v_{Nl} = 2.0 \pm 0.2$, and $r_{0s} = 274 \pm 7$ au. The fitting uses the Levenberg-Marquardt algorithm (Levenberg 1944; Marquardt 1963), implemented in the KaleidaGraph data analysis application. The best-fit column density ratio $v_{Nl}$ is consistent with the value $v_{Nl} = 2$ discussed in Section 2.6. Each of the density and column density values derived from these fit properties has corresponding uncertainty less than ~10% due to fitting.

The scale length $r_{0s}$ was corrected for beam smoothing by assuming that the observed column density profile $N(r)/N(0)$ is the convolution of the intrinsic profile and a Gaussian beam profile. The intrinsic profile can be approximated by a Gaussian of standard deviation $r_0$, for radii $r \lesssim r_0$, as indicated by equation (8) when $A = 1$. Then the intrinsic scale length is $r_0 = (r_{0s}^2 - \sigma_b^2)^{1/2}$. Assuming that the beam radius $\sigma_b = 140$ au is known with negligible error, the intrinsic scale length is recovered to be $r_0 = 238 \pm 7$ au and the column density scaling factor in dimensional units becomes $N = 3.5 \times 10^{22}$ cm$^{-2}$ $W(C^{18}O)[\text{K km s}^{-1}]^{-1}$.

The scale length and the observed lowest-contour radius are combined to give the dimensionless radius $\omega_l = r_l/r_0 = 19 \pm 1$ including uncertainties in $r_l$ and $r_0$. The observed and model mean column density are combined with the fit parameters and equation (32) to give the background, lowest-contour, and peak column densities $N_u$, $N_l$, and $N_0 = 0.5, 0.9,$ and $9 \times 10^{23}$ cm$^{-2}$ respectively. The factor of 2 ratio between $N_l$ and $N_u$ reflects the fit result $\nu_{Nl} = 2.0 \pm 0.2$. Each column density has relative uncertainty 0.1, dominated by the relative fit uncertainty in $\nu_{Nl}$.

## 5.2. Density Structure

The peak spheroid density $n_0$ is obtained by equating the observed and model mean column densities, and by substituting already determined values of $\omega_l$ and $r_0$, yielding $n_0 = (8 \pm 2) \times 10^7$ cm$^{-3}$. The relative uncertainty in $n_0$ and in the following density estimates is $\sigma_n/n \approx 0.2$, based on the uncertainties in the column density fit in Figure 2. The condensation formation efficiency is assumed to be $\epsilon_l = 0.3$ as discussed in Section 2.6. With $\nu_{Nl} = 2$ from the fit in Figure 2, this assumption sets the scale factor $\chi$ to be $\chi = 1$. Then $\nu_0 = \omega_l^2 = 360 \pm 20$, and $n_u = n_0/\nu_0 = 2 \times 10^5$ cm$^{-3}$. The densities associated with the lowest column density contour are obtained from equations (3) and (4), giving lowest-contour-density $n_l = 4 \times 10^5$ cm$^{-3}$ and mean density within $r_l$ equal to $\bar{n}_l = 8 \times 10^5$ cm$^{-3}$. The densities associated with the radius $r_1 = r_{\max,0} = 1080$ au discussed in Section 5.3.1 are the density at radius $r_1$, $n_1 = 4 \times 10^6$ cm$^{-3}$, and the mean density within $r_1$, $\bar{n}_1 = 8 \times 10^6$ cm$^{-3}$.

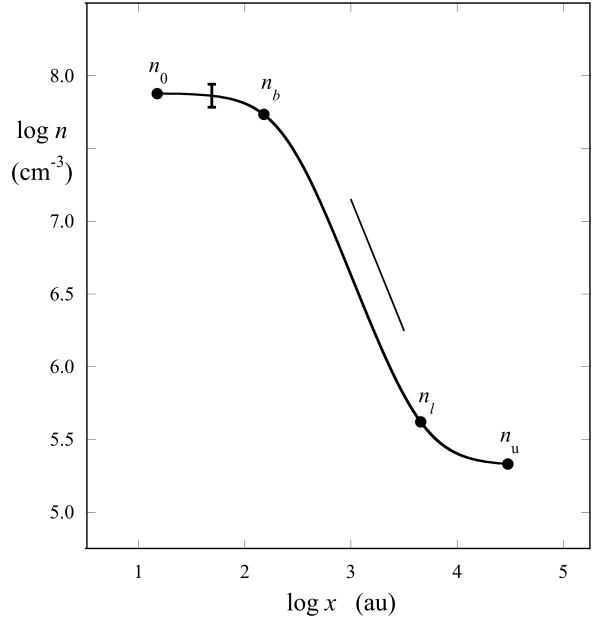

**Figure 3.** Density profile of the envelope of BHR71 IRS1. The profile is inferred from the observations in Figure 2 with the $p = 2$ Plummer model embedded in a uniform background, described in the text. *Filled circles* mark the peak density $n_0$, the background density $n_u$, and the range over which the profile is derived, from $n_b$ at the beam radius, to $n_l$ at the radius of the lowest closed contour. The error bar indicates 20% relative uncertainty. The *straight line* has the minimum slope of -1.80, close to the value -2 expected from the power-law of density with radius $n \propto r^{-2}$.

The continuous variation of the density with radius is shown in Figure 3, based on the foregoing source properties and on equation (3). The radius range extends outward beyond the observed envelope size scale, to $3 \times 10^4$ au, to show the limiting background value $n_u$ of the original medium. It extends inward, to 15 au, to show the limiting density as the radius decreases below one scale length in the Plummer model. Between the minimum and maximum radii, the density in Figure 3 approaches the dependence on radius $n \propto r^{-2}$ for the $p = 2$ Plummer density profile.

## 5.3. Field Direction Structure

### 5.3.1. Field Directions from $W(C^{18}O)$ Map

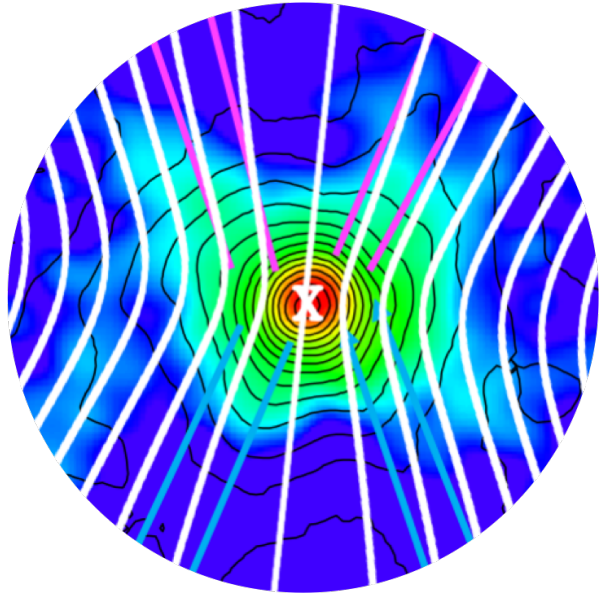

1000 au

**Figure 4**. SFF magnetic field lines in the plane of the sky (*white*) in the envelope of BHR71 IRS1, derived from the density model in Figure 3. The model parameters are density contrast $\nu_0 = 360$, scale length $r_0 = 238$ au, aspect ratio $A = 1$, magnetic axis inclination $i = 30°$ from the $z$-axis toward the observer, and plane-of-sky rotation of the $z$-axis $\theta_M = 6.4°$ W of N. These field lines are superposed on T19 Figure 18a, showing contours of $W(C^{18}O)$ increasing from $45\sigma$ in steps of $10\sigma$, where the rms noise $\sigma$ is $0.13$ K km s$^{-1}$ (*black*). *Red and blue lines* approximate inner outflow contours of $W(CO)$ in T19 Figure 18b. Model angles are illustrated in Figure 9.

Figure 4 shows magnetic field directions in the plane of the sky, estimated from the SFF density model parameters derived from the $W(C^{18}O)$ map analysis in Section 5.2. The inclination of the magnetic axis is equated to that of the outflow axis $i = 30°$ (T19). The plane-of-sky position angle is $\theta_M = 6.4°$ W of N, based on fitting to the polarization map as described in Section 5.3.2. This angle coincides with the outflow axis estimate of T19. The field directions were calculated as continuous flux tube curves as in Section 2.3 of Paper I. This display was chosen to help visualize the large-scale field pattern. In figure 4 some of the flux tubes (*white lines*) have opening angles similar to those of the outflow (*red and blue lines*).

**5.3.2. Field Directions from Polarization Map**

Attempts were made to estimate model parameters from the polarization data of H20, independent of those derived earlier from the $W(C^{18}O)$ map of T19. Three trials at fitting equation (21) to the polarization map were carried out. The data were masked to include positions between a fixed inner radius 300 au and a variable outer radius $r_{max}$. The inner positions were masked because their polarization directions clearly differ from the SFF model, possibly due to dust scattering (Wolf et al. 2008, Kataoka et al. 2015), rotation (Kataoka et al. 2015), and/or outflow motions. For simplicity this inner radius is denoted $r_{scat}$ although the enclosed polarization may have a more complex origin than scattering alone.

These fitting trials had limited success in matching the observed polarization pattern, and in determining useful parameter values. The parameters $v_0, r_0,$ and $i$ were poorly determined because they have significant correlation: an increase in hourglass pinch can arise from increasing $v_0$, increasing $i$, or from decreasing $r_0$. This property contrasts with the column density fitting in Section 5.1, which gives better-determined parameter values. The two types of fit may differ because the column density structure is much simpler than the polarization structure, and because the column density fits do not need to mask data within the central few hundred au. This region is important for setting the scale length of the density and column density structure.

The most successful polarization map fits treat the magnetic axis angle $\theta_M$ as a single free parameter, with remaining parameters fixed. The fixed parameters were determined from the column density fitting ($r_0$ and $v_0$) or assumed based on other observational information ($A, i,$ and $\alpha$). In BHR71 IRS1 the best-fit angle was found to be $\theta_M = 6.4°$ W of N with uncertainty 0.2° over the range 660 au $\leq r_{max} \leq$ 1400 au.

The successful fitting of the projected magnetic axis direction $\theta_M$ from the polarization map contrasts with the column density map fitting in Section 5.1, which cannot determine $\theta_M$ well for spheroids of low aspect ratio. Angle fitting to determine $\theta_M$ allows comparison between the magnetic and outflow axis angles, to measure the degree of misalignment. Axis misalignment may be important in models of disk formation (Matsumoto & Tomisaka 2004, Li et al. 2013, Hull et al. 2013, Hull & Zhang 2019).

### 5.3.3. Combining Column Density and Polarization Map Data

The foregoing results suggest that column density analysis should be done as a first pass, to set nominal parameter values of $v_0, r_0,$ and $A$. It may be possible to also obtain inclination $i$ from a column density map if it has the elliptical projected shape of an inclined oblate spheroid. Assuming $\theta_M$ is along the poles of the star, the inclination may also be estimated via independent disk observations, or from independent outflow observations.

### 5.3.4. Comparison of Model Field Lines with Polarization Map

The model fit to the observed polarization map (H20) was evaluated to determine the map region where the model fits best, using the radius values $r_{scat}$ to $r_{max}$ as described above. Over this range equation (21) was used to find the best-fit angle $\theta_M$ of the projected magnetic axis, assuming values of $A, r_0, v_0,$ and $i$ given earlier. Then the model field angle $\theta_B$ was calculated at each position to give the angle difference $\Delta\theta = \theta_{B,pol} - \theta_B$, the distribution of $\Delta\theta$ values, and the standard deviation $\sigma_{\Delta\theta}$ of the distribution. This process was repeated for increasing values of $r_{max}$. The resulting variation of $\theta_M$ and of $\sigma_{\Delta\theta}$ with $r_{max}$ is shown in Figure 5.

Figure 5 is labelled with four zones of outer radius $r_{max}$. These zones are based on the angle properties $\theta_M$ and $\sigma_{\Delta\theta}$ evaluated over the map radius range $r_{scat} \lesssim r \leq r_{max}$, and on the morphological appearance of the polarization map. When $r_{max}$ lies within $r_{scat}$ no magnetic fits are possible, as discussed above. For $r_{scat} \lesssim r \leq r_{max} = 660$ au, fits to determine $\theta_M$ either fail, or their best-fit values of $\theta_M$ vary significantly from one value of $r_{max}$ to the next. These unstable fits probably occur because the radius range from 300 to 660 au encloses too few independent measurements for useful statistics.

As $r_{max}$ increases from 660 au to 1400 au, the best-fit angles $\theta_M$ and the standard deviations $\sigma_{\Delta\theta}$ for $r_{scat} \lesssim r \leq r_{max}$ become stable, and the polarization pattern has a well-defined hourglass shape. The magnetic axis angle has the same best-fit value $\theta_M = 6.4°$ for each value of $r_{max}$, with median one-sigma fit error $0.2°$. This magnetic axis angle agrees closely with the outflow axis angle $6°$ estimated from the bisector of the CO outflow contours (T19). The values of $\sigma_{\Delta\theta}$ have mean $17.6°$ and standard deviation $0.7°$. The standard deviation $\sigma_{\Delta\theta}$ has its minimum value $\sigma_{\Delta\theta,0} \equiv 17.3°$ when $r_{max} = r_{max,0} \equiv 1080$ au. Thus $r_{scat} \lesssim r \leq r_{max,0}$ is the radius range with the closest match of the SFF model angles $\theta_B$ to the polarization directions $\theta_{B,pol}$. The value of $r_{max,0}$ is adopted for display of the magnetic field lines in Figure 6 and for evaluation of the DCF mean magnetic field strength in Section 5.4.

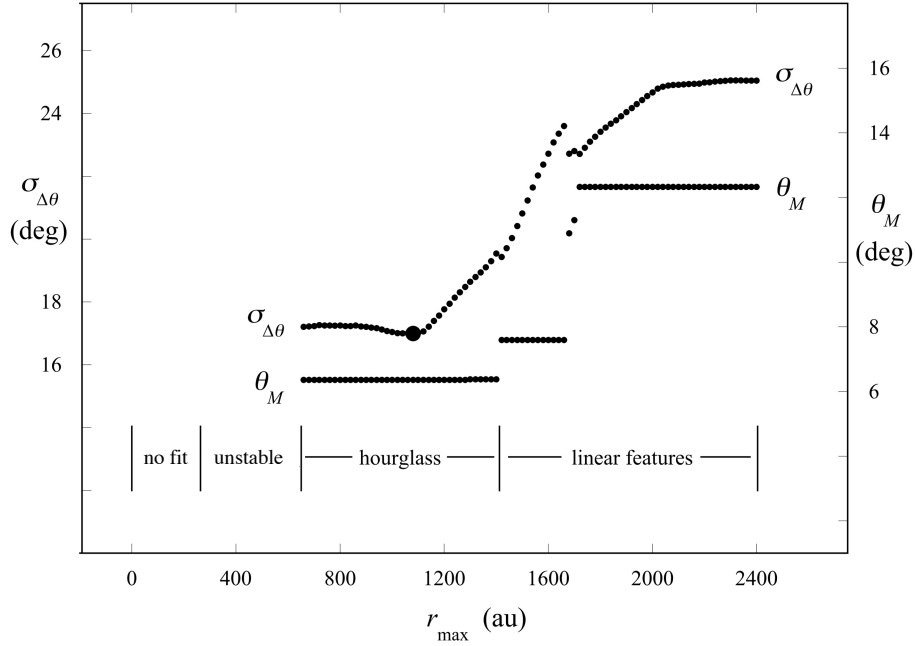

**Figure 5.** Best-fit value of the projected magnetic axis angle $\theta_M$ and the standard deviation $\sigma_{\Delta\theta}$ of the distribution of angle differences $\Delta\theta = \theta_{B,\text{pol}} - \theta_B$ in the envelope of BHR71 IRS1, for radii $r$ between $r_{\text{scat}} = 300$ au and an increasing outer radius $r_{\text{max}}$. The angle $\theta_M$ increases from North (N) to West (W). The large filled circle marks the radius $r_{\text{max},0} = 1080$ au where $\sigma_{\Delta\theta}$ has a local minimum of 17.3° for radii $r_{\text{scat}} \leq r \leq r_{\text{max},0}$. The map radii are divided into four zones discussed in the text.

As $r_{\text{max}}$ increases beyond 1400 au, $\sigma_{\Delta\theta}$ increases more steeply with $r_{\text{max}}$, and the polarization pattern departs from a map-filling hourglass. Instead it is dominated by linear features having nearly uniform direction within each feature. These features may correspond to accretion flows (H20). A small spike at $r_{\text{max}} \approx 1700$ au coincides with the onset of a polarization feature associated with IRS2. At $r_{\text{max}} \approx 2400$ au, $\sigma_{\Delta\theta}$ reaches its maximum value of ~25°.

Figure 5 shows that $\sigma_{\Delta\theta}$ can vary significantly, depending on the size of the region where the hourglass pattern dominates, and on the size of the region over which $\sigma_{\Delta\theta}$ is evaluated. This variation has an equally significant effect on the mean field strength estimated by the DCF method, which varies as $1/\sigma_{\Delta\theta}$. If $r_{\max}$ is larger than the region where an hourglass model applies, $\sigma_{\Delta\theta}$ will be overestimated and the mean field strength will be underestimated.

Following the results of Figure 5, Figure 6 shows the SFF field direction pattern in Figure 4, oriented to $\theta_M = 6°$, restricted to radii within 1080 au, and superposed on the 1.3 mm ALMA polarization map (H20). The polarization directions $\theta_{B,\text{pol}}$ have been rotated through 90° to show the corresponding magnetic field directions. Figure 6 shows that the model and map directions have good qualitative agreement. They agree more closely in the radius range 300 au $\leq r_{\max} \leq$ 1080 au than at smaller or larger radii, as expected from Figure 5.

The analysis of standard deviations $\sigma_{\Delta\theta}$ in Figure 5 is useful to select the map region of best model fit. The underlying distributions of $\Delta\theta$ give additional insight into the quality of the SFF model fits and into their variation from small to large map radius. For this purpose Figure 7 shows three histograms plotted for the same range of angles.

In Figure 7, the distribution of the difference between observed and model angles within $r_{\max,0} = 1080$ au in (b) is distinctly narrower than the distributions in (a) and (c). In (a) a constant angle value has been subtracted from the observed distribution to center the plot. The reduction in width from (a) to (b) indicates that an appropriate SFF model accounts for the bimodal structure of the hourglass distribution, and for some of its intrinsic width. The reduction in width from (c) to (b) reflects the trend of increasing $\sigma_{\Delta\theta}$ with $r_{\max}$ in Figure 5. It shows that this trend is due to increasing positive and negative deviations $\Delta\theta$ with increasing radius, with an excess of positive over negative deviations.

The distribution in (b) has $\sigma_{\Delta\theta} = 17.3°$, dominated by the asymmetrical wing on the positive side of the mode. The standard deviation differs by a factor ~2 between the angles on the negative side of the mode (7.5°) and on the positive side of the mode (14.2°). The spatial origin of the positive wing is limited to two distinct zones where $\theta_{B,\text{pol}}$ exceeds $\theta_B$ by 10°- 40°, one to the ESE of the center and one to the WSW of the center. These zones have ordered polarization directions which resemble neither hourglass structure nor random turbulent fluctuations described by the DCF model. Nonetheless they are counted as contributing to the magnetic field estimate, following standard practice in DCF analysis.

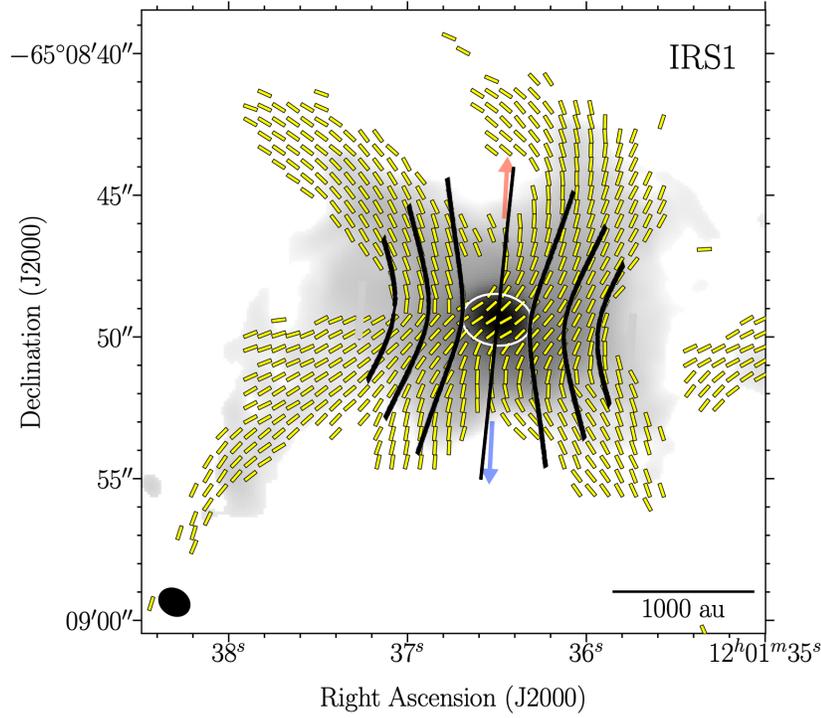

**Figure 6**. 1.3 mm ALMA map of polarization directions and SFF model magnetic field lines in the BHR71 IRS1 region. Each observed polarization segment (*yellow*) has been rotated through 90° to represent polarization due to magnetically aligned grains (H20 Figure 2). The white ellipse encloses the region where the polarization is assumed to be due to dust scattering and other processes not included in the SFF model. The curves (*black*) show magnetic field lines predicted by the SFF model as in Figure 4, within the radius $r_{\max} \approx 1080$ au which gives the minimum value of $\sigma_{\Delta\theta} = 17.3°$.

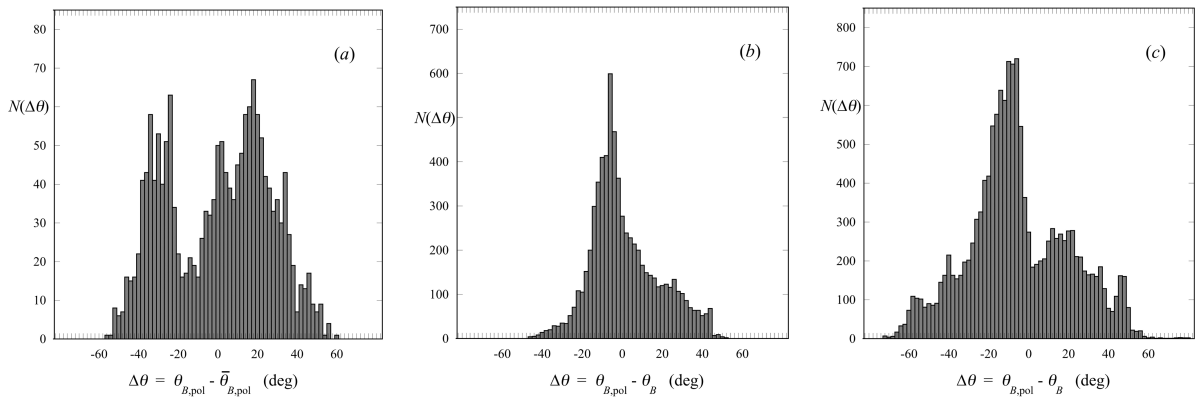

**Figure 7.** Distributions of angle difference $\Delta\theta$ in the envelope of BHR71 IRS1, for different models of magnetic field direction and different ranges of map radius $r$. In each panel, $\Delta\theta$ is the difference between the observed polarization angle $\theta_{B,\text{pol}}$, rotated by 90° (H20), and one of three models $\theta_{\text{mod}}$: (a) $\theta_{\text{mod}}$ = a uniform angle, equal to the mean map angle $\bar{\theta}_{B,\text{pol}}$, within $r_{\text{max},0} = 1080$ au; (b), $\theta_{\text{mod}}$ = the SFF model $\theta_B$, within $r_{\text{max},0}$, and (c) $\theta_{\text{mod}}$ = the SFF model $\theta_B$, within $2r_{\text{max},0}$. Each angle $\theta_B$ and $\theta_{B,\text{pol}}$ increases from N to W.

### 5.4. Field Strength Properties
### 5.4.1. Mean Field Strength and Magnetically Critical Mass

The DCF mean field strength was calculated by evaluating equation (26) in the best-fit range of map radii $660$ au $\leq r \leq 1080$ au determined from Figure 5. The mean nonthermal velocity dispersion $\bar{\sigma}_{NT} = 0.17$ km s$^{-1}$ was taken from $\overline{\Delta v}_{\text{FWHM}} = 0.4$ km s$^{-1}$, the mean FWHM line width of C$^{18}$O in T19 Figure 18, assuming optically thin emission and Gaussian line shape. The uncertainty in mean field strength was calculated from standard propagation of errors in $Q_B, \bar{n}$, and $\sigma_{\Delta\theta}$. The resulting mean ± standard deviation is $\langle\bar{B}\rangle = 0.7 \pm 0.2$ mG. If the field axis inclination is based on the outflow direction rather than on the statistical average in equation (25), this DCF field strength estimate is reduced by ~8%, which is negligible. The relative uncertainty ~0.2 in $\langle\bar{B}\rangle$ also dominates the uncertainty in the SFF field strength values obtained by scaling $\langle\bar{B}\rangle$ with density and mean density in Section 5.4.2.

The ratio of mass to magnetically critical mass was evaluated from equation (27). The mean of $\bar{n}r/\langle\bar{B}\rangle$ was calculated over the same range of $r$ as for the mean field strength above. The uncertainty is based on similar propagation of errors as for $\langle\bar{B}\rangle$, yielding $M/M_c = 1.5 \pm 0.4$. Magnetic forces were evidently too weak to prevent formation of the protostar IRS1. Since at present $M/M_c > 1$, magnetic forces also appear too weak to prevent further star-forming collapse.

The conclusion $M/M_c > 1$ follows from the DCF analysis, based on the assumption that Alfvénic fluctuations are associated with polarization dispersion and nonthermal line widths. This conclusion is consistent with the independent SFF assumption that $M/M_c > 1$, as discussed in Section 3.2. This consistency justifies the use of $\langle\bar{B}\rangle$ as a reference field for the calculation of the field strength profile in the next section.

### 5.4.2. Field Strength Structure

The mean field strength $\langle \bar{B}_1 \rangle = 0.7$ mG in Section 5.4.1 is combined with the density parameters and equations (28)-(31) to give the field strength profile along the x-axis as given in Section 3.3. Since the envelope model is spherical, $A = 1$ and $\omega = \xi$ in equations (28)-(31).

The field strength profile is shown in Figure 8. It has background value $B_u = 60$ µG, values 0.3 mG to 0.6 mG in the hourglass polarization zone where the SFF model has best match to the polarization map, and peak value 3 mG. The magnetic field strength profile has similar shape to the density profile in Figure 3. The field strength approaches the power-law dependence on radius $B \sim r^{-4/3}$. This relation is expected from the dependence of field strength on density as $B \sim n^{2/3}$ for weak-field flux-freezing (M66, Crutcher 2012, Mocz et al. 2017), and from the $p = 2$ Plummer dependence of density on radius as $n \sim r^{-2}$ discussed in section 5.2.

The field strength in Figure 8 has constant direction $\widehat{\boldsymbol{B}} = \hat{\boldsymbol{z}} \cos i - \hat{\boldsymbol{y}} \sin i$, parallel to the magnetic axis. Here the z-axis lies in the plane of the sky at best-fit angle $\theta_M = 6.4°$ W of N, in good agreement with the estimated direction of the outflow axis in the plane of the sky (T19). The magnetic axis is inclined from the z-axis toward the observer by $i = 30°$, assumed to match the estimated outflow inclination (T19). In the angle diagram in Figure 9, the magnetic axis coincides with the vector $\vec{r}$ when $\theta = 0$.

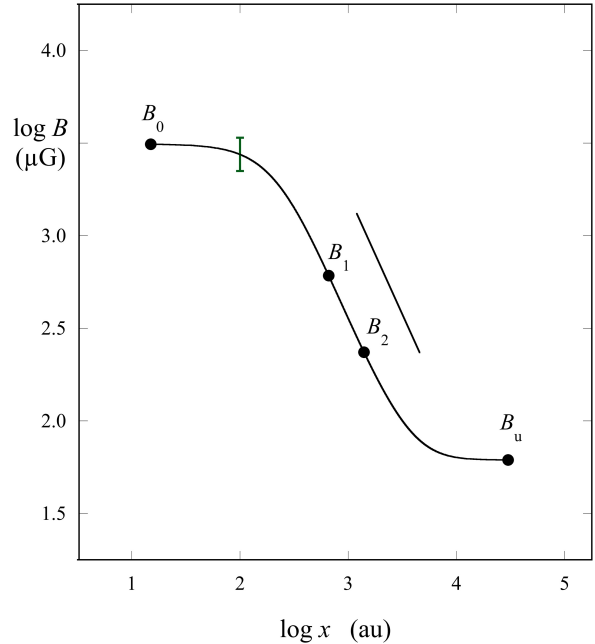

**Figure 8.** Profile of magnetic field strength in the envelope of BHR71 IRS1, as a function of radius along the $x$ - axis in the $z = 0$ plane. The field strength is inferred from the density model in Section 2, from the velocity dispersion in T19, and from the polarization observations of H20 in Figure 6. The error bar indicates typical relative uncertainty of $\sim 0.2$. *Filled circles* mark the peak field strength $B_0$, the background field strength $B_u$, and the range of field strength $B_1$ to $B_2$ in the best-fit hourglass polarization zone. The *straight line* has the minimum slope -1.28, close to the value -4/3 expected from the power-laws of field strength with density $B \propto n^{2/3}$ and density with radius $n \propto r^{-2}$.

The DCF field strength estimate of 0.7 mG in BHR 71 IRS1 is intermediate in the range of DCF estimates of mean field strength in star-forming regions. Among starless cores and globules, the DCF field strength is 7 μG in CB81 (Kandori et al. 2020b), 26 μG in B68 (Kandori et al. 2020c), 29 μG in FeSt 1-457 (Kandori et al. 2018), and 30 μG in B335 (Kandori et al 2020d). Among low-mass star-forming regions, the DCF field strength is 3 mG in the star-forming cloud L1157 (Stephens et al. 2013) and 5 mG in the region NGC1333 IRS4 forming a small group (Girart et al. 2006). In the central part of the cluster-forming clump G31+0.31, the field strength is 8-13 mG (Beltrán et al. 2019). This ranking of relative mean field strengths seems plausible in relation to the typical column densities in each region, and to the stellar masses being formed in each region. However, differences in resolution and technique from one study to the next add significant uncertainty.

## 6. DISCUSSION
### 6.1. Summary

This paper extends the SFF model of Paper 1 to provide improved estimates of magnetic field strength and structure in a spheroidal molecular cloud. The estimates are based on observations sensitive to cloud column density, polarization of magnetically aligned grains, and nonthermal velocity dispersion. The model assumes conservation of mass, shape, and flux during the formation of a centrally condensed $p = 2$ Plummer spheroid in a uniform medium. Sections 2 and 3 describe the equations and procedures used. They provide column density models of spherical, prolate, and oblate Plummer spheroids having any aspect ratio and inclination. These models are matched to observed maps of optically thin dust or molecular line emission, to estimate the parameters of the appropriate spheroid density model: its peak density $n_0$, background density $n_u$, scale length $r_0$, and aspect ratio $A$.

The spheroid density model is combined with estimates of the magnetic axis inclination and rotation on the plane of the sky to predict the hourglass shape and orientation of the plane-of-the-sky magnetic field lines. The axis inclination $i$ is assumed, or it is approximated to equal the inclination of the associated outflow, if known. The axis rotation $\theta$ on the plane of the sky is obtained by matching the estimated outflow axis rotation, or by fitting the field direction model to the observed polarization map.

The resulting field direction model is compared to the observed polarization map to find the radius within which the model and map have the best match. At this radius $r_1$ the distribution of angle differences $\Delta\theta \equiv \theta_{B,\text{pol}} - \theta_B$ has the smallest standard deviation $\sigma_{\Delta\theta,0}$. This standard deviation is used to estimate the mean field strength within $r_1$.

The mean field strength within $r_1$ has plane-of-sky component $(\bar{B}_1)_p$ estimated via DCF analysis, from $\sigma_{\Delta\theta,0}$, the model mean density $\bar{n}_1$, and the observed mean nonthermal velocity dispersion $(\bar{\sigma}_{NT})_1$. The total field strength corresponding to $(\bar{B}_1)_p$ has expectation value $\langle \bar{B}_1 \rangle$ obtained by averaging over inclinations, which are assumed equally likely.

The mean field strength $\langle \bar{B}_1 \rangle$ sets the magnitude of the continuous profile of field strength with radius given in equation (28). The profile is derived from the SFF dependence of field strength on density and mean density at each radius, which varies approximately as $B \propto n(\bar{n})^{-1/3}$. The field strength profile $B(r)$ is derived from the column density map, so it has finer resolution than the mean field strength within $r_1$ by the factor $r_1/r_b$ where $r_b$ is the beam radius.

Section 4 summarizes in step-by-step form the application of the models in Sections 2 and 3 to estimate the magnetic field structure from observed maps.

Section 5 applies these steps to ALMA observations of the envelope associated with the protostar BHR71 IRS1. The observations were made with ~300 au resolution in the integrated intensity of $J = 2\text{-}1$ $C^{18}O$ (T19) and in 1.3 mm continuum polarization (H20). The density model obtained from the $C^{18}O$ map ranges from background density $n_u = 2.1 \times 10^5$ cm$^{-3}$ to peak density $n_0 = 7.5 \times 10^7$ cm$^{-3}$, with scale length $r_0 = 240$ au. The density model generates the shape of the field direction model. Fitting the field direction model to the polarization map sets the magnetic axis orientation in the plane of the sky to be $\theta = 6$ deg W of N. The resulting distribution of polarization - field angle differences has minimum dispersion $\sigma_{\Delta\theta} = 17.0$ deg within map radius 1080 au.

Applying DCF analysis gives a mean field strength $\langle \bar{B}_1 \rangle = 0.7$ mG with relative uncertainty $\approx 0.2$, and a corresponding ratio of mass to magnetically critical mass $M_l/M_{cl} = 1.5 \pm 0.4$. This ratio indicates that the envelope field energy is too weak to prevent further gravitational collapse of the envelope mass. The continuous variation of field strength with equatorial radius ranges from the background field strength $B_u = 60$ µG to the peak field strength $B_0 = 3$ mG, also with relative uncertainty $\approx 0.2$.

## 6.2. Significance

These results give for the first time detailed estimates of magnetic field strength and structure in an ideal-MHD flux-freezing model of a spheroidal condensation, based on observations sensitive to column density, polarization of magnetically aligned grains, and nonthermal velocity dispersion. In the example of the envelope of BHR71 IRS 1, the resolution ~300 au is the resolution of the column density map. It is significantly finer than the resolution ~2200 au of the map-average field strength obtained from DCF analysis. This section compares the SFF estimates presented here with other recent field strength estimates.

### 6.2.1. Comparison with Other Field Strength Estimates from Polarization Maps

The present models use more observable information than many previous field estimation models. They infer field strength structure from an observed map sensitive to column density as well as from an observed polarization map. The column density map can strongly constrain the polarization model, since the column density map generates the parameters of the spheroid density model, which largely specifies the model field directions to compare with the polarization map.

Numerous studies estimate field strength from well-resolved polarization maps. Some estimate field strength by assuming that the tension of a curved field line is in force balance with the gravitational pull toward the center of mass (Schleuning 1998, Li et al. 2015; see also Koch et al. 2012, 2013, 2018). However, most studies use the DCF method as discussed below.

### 6.2.2. Comparison with Other Field Strength Estimates Using DCF Analysis

This paper estimates field strength structure in a Plummer spheroid by multiplying the mean DCF field within dimensionless radius $\omega_1$ by the ratio of SFF field strength to SFF mean field strength, i.e. $B(\omega) = \langle \bar{B}(\omega_1) \rangle_{DCF} [B(\omega)/\bar{B}(\omega_1)]_{SFF}$ as in equation (28). In this estimate the DCF field strength is a coarse-resolution average over the polarization map. It sets the field strength scale for the finer-resolution SFF estimate of field strength structure. This section describes differences between the present method and other DCF studies.

Field strength estimates based on DCF analysis differ from each other mainly in how the ordered polarization is removed to obtain $\sigma_{\Delta\theta}$, the dispersion in directions attributed to Alfvénic fluctuations. Some studies use polarization structure predicted by models assuming fixed mass-to-flux ratio (Frau et al. 2011, Alves et al. 2018, Beltrán et al. 2019). Others estimate the background structure to be removed by first locally smoothing the polarization map (Pattle et al. 2017, Kwon et al. 2019). Several studies represent the ordered polarization with nested parabolas (Girart et al 2006, Rao et al. 2009, Stephens et al. 2013, Qiu et al. 2014, Kandori et al. 2017, 2018). A comparison of results between using nested parabolas and the SFF structure is given in Kandori et al. 2020c). Among these papers, Frau et al. (2011) compare model Stokes *I* maps with observations, but they choose their best-fit models only from the polarization maps. None of these foregoing estimates match both the column density and polarization maps as in the present paper.

The field strength estimates in this paper have three features which seem more realistic than some other estimates of field strength based on DCF analysis.

(1) Better-constrained ordered component. In the SFF model the field angle $\theta_B$ subtracted from $\theta_{B,pol}$ at each map point to obtain $\sigma_{\Delta\theta}$ and $\langle\bar{B}_1\rangle$ is based on observed column densities and a physical flux-freezing model. In contrast, estimates of the ordered polarization component based on polarization map smoothing, or on simple analytic functions such as parabolas, ignore the constraints on field structure due to density structure and flux freezing.

(2) Finer spatial resolution. The spatial resolution of a SFF estimate of field strength is essentially the resolution of the column density map on which it is based. This resolution can be substantially finer than that of the DCF estimate of mean field strength, since the DCF estimate requires at least enough independent angle measurements to obtain a statistically significant estimate of $\sigma_{\Delta\theta}$. If a column density map and a polarization map have the same beam radius $r_b$ and if $\sigma_{\Delta\theta}$ is estimated over all the map positions within $r_1$, the SFF resolution advantage is a factor $\sim r_1/r_b$ as noted in Section 6.1. For the $W(C^{18}O)$ maps of BHR 71 IRS1 in Figures 2 and 4, this ratio is a factor $\sim 7$. This resolution advantage over DCF analysis is also a feature of the Core Field Structure method of A19.

This resolution advantage can be especially important in estimating the peak field strength in a condensation, since the peak can be significantly greater than the mean. In BHR71 IRS1 the mean field strength is 0.7 mG due to the DCF estimate with effective resolution 2200 au, while the mean field strength centered on the same position is 2.8 mG due to the SFF estimate with effective resolution 300 au. In this example the DCF field strength underestimates the SFF field strength by a factor ~4.

(3) Field strengths in nonspherical condensations. The SFF model extends the range of cloud geometries which can be analyzed, from spherical envelopes, cores, and clumps to filaments and filamentary cores, which can be modelled as prolate spheroids, and to flattened envelopes and compressed layers, which can be modelled as oblate spheroids. SFF analysis allows estimation of how the field strength varies along axes having different scale lengths, with finer resolution than with DCF analysis alone. The example of BHR71 IRS1 is conveniently approximated here by a spherical model, but the estimates of column density, density, field directions and field strength can be easily extended to oblate and prolate models using the procedures in Sections 2 and 3.

### 6.2.3. Comparison with Auddy et al. (2019; A19)

This paper extends the "Core Field Structure" (CFS) models and procedures in A19. Each paper relies on the DCF relation between the nonthermal component of a suitable spectral line width and Alfvénic fluctuations, to set the magnitude scale of the radial profile of field strength in a condensation. In each paper this profile increases with density, according to flux freezing and a density model derived from observations sensitive to column density. Each paper has the same resolution advantage over the DCF estimation of mean field strength. The main difference is that this paper estimates the reference field strength from polarization fitting, while A19 assumes a fixed level of Alfvén wave excitation at the transonic radius of a dense core.

### 6.3. Limitations and Prospects
### 6.3.1. Assumptions and Uncertainties

The SFF model has limitations discussed in Paper 1 due to its idealized assumptions that a condensation arises from an original medium with uniform density and magnetic field, while conserving mass, flux, and spheroidal shape. Unrealistic features of the model are its static initial and final states, and its assumptions that field lines are not significantly affected by ordered or turbulent flows, by magnetic tension forces, and by nonideal MHD processes such as ambipolar diffusion.

The SFF model assumes that the observed column density structure has a detectable background. The column density gradient with radius should make a smooth transition from steep to shallow, and the observations should have high enough signal-to-noise ratio to detect and fit this transition, as in Figure 2. There the background fit error is less than 10%, and it has a negligible effect on the total error estimate. If no such background can be reliably identified, estimation of the density contrast $v_0$ by methods described in Paper I will be more uncertain. If this uncertainty is larger than a factor ~3 it will become the primary source of uncertainty in field strength.

The DCF coefficient $Q_{B_p}$ in Section 3.1 relates the mean field strength in a condensation to its mean density and velocity dispersion. OSG01 obtained $Q_{B_p} = 0.5$ for simulated condensations whose structure is dominated by turbulent motions. In Section 3.1 it is assumed that this coefficient is also appropriate to estimate the mean field strength in a centrally condensed Plummer spheroid. This assumption has not been tested by a detailed calculation or simulation. It is assumed here that the coefficient for a Plummer spheroid departs from $Q_{B_p} = 0.5$ by less than the 20% relative uncertainty estimated in Section 3.1.

The dispersion of angle differences $\sigma_{\Delta\theta}$ in the application of this model in Paper I to observations of VLA 1623A (Sadavoy et al. 2018) and to BHR71 IRS1 in this paper, is in each case less than ~20 deg. It is therefore comparable to dispersions reported in other comparisons of physical models to observations of polarization due to grains aligned with the magnetic field (Goncalves et al. 2008, Alves et al. 2018, Beltrán et al. 2019).

### 6.3.2. Possible Applications

A limitation noted in Paper 1 is the assumption of purely poloidal fields, i.e. no toroidal component due to rotational dragging and stretching of field lines. This concern has been addressed in this paper with a semi-analytic representation of toroidal field line structure, in Section 2.8.4 and illustrated in Figure 1(d). It may be useful to apply this representation to recent observations of L1448 IRS 2, whose ALMA polarization pattern shows a strong toroidal component (Kwon et al. 2019).

The density models in Paper 1 and in this paper are based on the embedded $p = 2$ Plummer function, due to its wide use in modelling star-forming filaments and cores. However the basic procedures in these papers can be applied to any symmetric function of radius having a single local maximum. It remains for the future to test how much field patterns vary when density functions have different dependence on radius than that of the $p = 2$ Plummer function.

The prevalence of field line patterns having evidence of ordered flows (e.g. Sadavoy et al. 2019, Hull & Zhang 2019, H20) suggests that it may be useful to combine the present static SFF models with a component of dynamical flows, to improve the match of model to observed polarization patterns.

## 7. CONCLUSION

The main features of this paper are:

1. Models and procedures are described to estimate the structure of magnetic field strength in a spheroidal star-forming condensation, with finer resolution than the Davis-Chandrasekhar-Fermi (DCF) analysis of Alfvénic fluctuations. The method combines features of the DCF model and the Spheroid Flux Freezing (SFF) model of flux and mass conservation in Paper 1.

2. The method analyzes a line or continuum map sensitive to column density and a polarization map sensitive to magnetic field structure. A Plummer-spheroid model of column density is matched to the column density map to give the spheroid density structure. SFF analysis of this structure gives a spatial model of its associated field directions. This model is fit to the polarization map to give the dispersion $\sigma_{\Delta\theta}$ between map and model angles.

3. DCF analysis of $\sigma_{\Delta\theta}$ with the mean condensation density and nonthermal velocity dispersion gives the mean field strength within the radius where the SFF model has the best match to the observed polarization directions. SFF analysis of the spheroid density structure gives the ratio of local to the DCF field strength. Combining these results gives the structure of the magnetic field strength, with the resolution of the column density map.

4. This paper provides the analytic basis of the models and a step-by-step guide to their application. It applies these steps to ALMA observations of dust continuum and molecular line emission, and to polarized continuum emission at 1.3 mm wavelength, from the envelope of the protostar BHR 71 IRS-1 with resolution ~300 au (Bourke et al. 1995, T19, H20).

The main findings of this paper are:

1. A spherical $p=2$ Plummer column density model in a uniform background was matched to the ALMA map of $J = 2 - 1$ $C^{18}O$ integrated intensity in the envelope of BHR IRS1 (T19). This gives column density ranging from background value $5 \times 10^{22}$ cm$^{-2}$ to a peak value $9 \times 10^{23}$ cm$^{-2}$ with scale length 240 au. The lowest closed contour of column density encloses 2.0 $M_\odot$ in a spherically symmetric model.

2. Analysis of the column density structure gives the envelope density model, ranging from a background value $2 \times 10^5$ cm$^{-3}$ to peak value $8 \times 10^7$ cm$^{-3}$. This density model predicts the plane-of-sky field direction model for comparison with the polarization map of H20.

3. The field direction model fits the polarization map best within ~1100 au, giving $\sigma_{\Delta\theta} = 17$ deg, with mean DCF field strength ~700 µG, with relative uncertainty ~ 0.2. Scaling this field strength with the SFF dependence on density gives the magnetic field structure in the equatorial plane, ranging from a background value 60 µG to peak value 3 mG. The field strength profile has a resolution advantage over the mean DCF field strength of a factor ~7. The peak field strength exceeds the mean DCF field strength by a factor ~5. The ratio of envelope mass to magnetic critical mass is $1.5 \pm 0.4$, indicating that the envelope mass is supercritical.

**Acknowledgements**   The authors thank Maite Beltrán, Thushara Pillai, and Sarah Sadavoy for helpful comments, and Ana Georgescu for assistance with fitting model column density and field directions. The authors thank the referee, whose constructive comments improved the paper. C.L.H.H. acknowledges the support of both the NAOJ Fellowship as well as JSPS KAKENHI grant 1K13586.

# APPENDIX A

Figure 9 illustrates the model coordinate system and angles defined in Section 2.

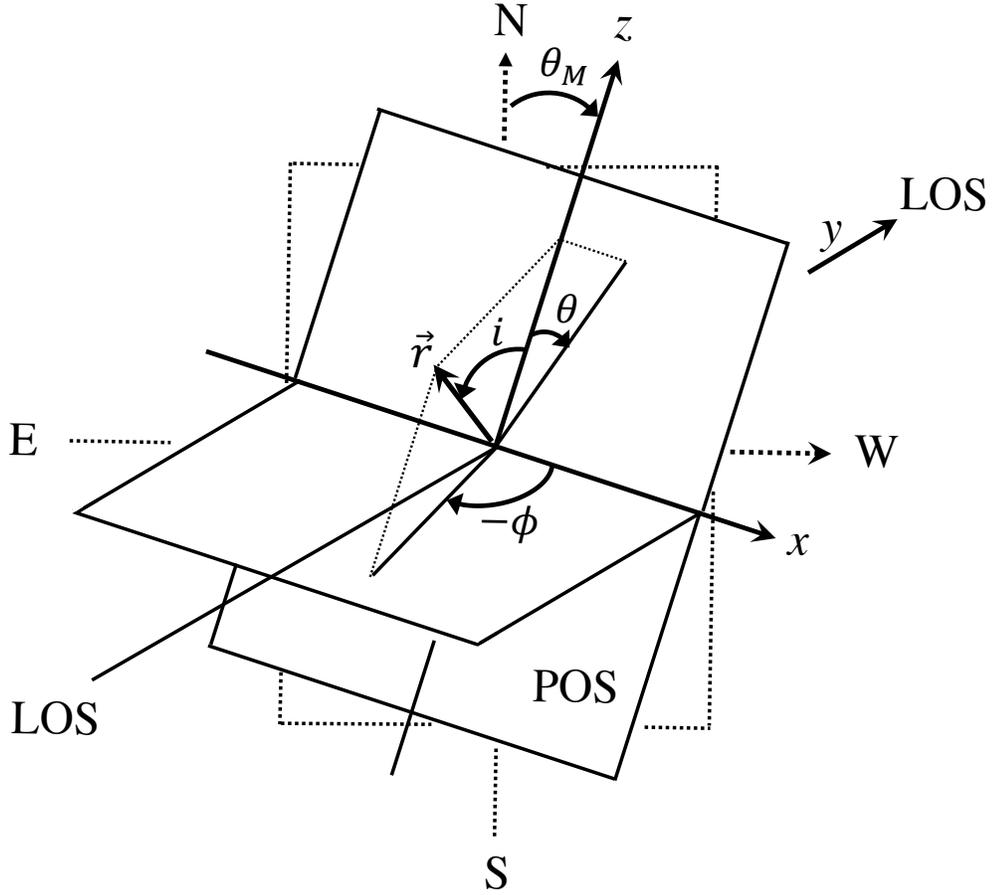

**Figure 9.** Geometry of model directions and angles. The *y*-axis lies along the line of sight (LOS). The *y*-axis is common to the celestial (dotted) and magnetic (solid) coordinate planes, which are centered on the column density peak and which lie in the plane of the sky (POS). The *z*-axis coincides with the polarization map symmetry axis, which is the projection of the magnetic axis onto the POS. The magnetic plane (*solid lines*) is rotated clockwise about the *y*-axis by $\theta_M$ from the celestial plane (*dotted lines*). A radius vector $\mathbf{r}$ is inclined from the *z*-axis through angle $i$ in the *r-z* plane. Its projection onto the *x-z* plane lies at angle $\theta$ from the *z* - axis, and its projection onto the *x-y* plane lies at azimuth angle $\phi$ from the x-axis, where $\phi = \cos^{-1}(\tan\theta/\tan i)$ for $|\tan\theta/\tan i| \leq 1$.

# APPENDIX B

Table 3 summarizes the main symbols used in this paper.

**Table 3**
Principal Symbols Used in This Work

| (1) Symbol | (2) Meaning | (3) Section | (4) Equation | (5) Figure | (6) Table |
|---|---|---|---|---|---|
| $B$ | magnetic field strength | 1 | 28 | 8 | |
| $n$ | gas volume density | 1 | | 3 | |
| $\kappa$ | exponent in $B \propto n^\kappa$ | 1 | | | |
| $x$ | horizontal coordinate | 2.1 | 1 | | |
| $y$ | LOS coordinate [a] | 2.1 | | | |
| $z$ | vertical coordinate | 2.1 | 2 | | |
| $\theta_M$ | angle from N to $z$-axis | 2.1 | 1 | 9 | |
| $\theta$ | angle from $z$-axis toward $x$-axis | 2.1 | | 9 | |
| $r$ | radius vector | 2.1 | | 9 | |
| $i$ | angle from $z$-axis to $r$ | 2.1 | | 9 | |
| $\phi$ | angle from $x$-axis toward $y$-axis | 2.1 | | 9 | |
| $p$ | exponent in Plummer $n$-profile | 2.1 | | | |
| $x_c$ | horizontal celestial coordinate | 2.2 | 1 | | |
| $z_c$ | vertical celestial coordinate | 2.2 | 1 | | |
| $x_{c0}$ | $x_c$ at column density map peak | 2.2 | 1 | | |
| $z_{c0}$ | $z_c$ at column density map peak | 2.2 | 1 | | |
| $n_u$ | density of uniform background gas | 2.3 | | | |
| $\nu$ | normalized density $n/n_u$ | 2.3 | 3 | | |
| $n_0$ | peak density above background | 2.3 | | | |
| $\nu_0$ | density contrast ratio $n_0/n_u$ | 2.3 | 3 | | |
| $\omega$ | normalized spheroid radius | 2.3 | 3 | | |
| $r_0$ | Plummer density profile scale length | 2.3 | | | |
| $\xi$ | normalized horizontal coordinate $x/r_0$ | 2.3 | | | |
| $\eta$ | normalized LOS[a] coordinate $y/r_0$ | 2.3 | | | |
| $\zeta$ | normalized vertical coordinate $z/r_0$ | 2.3 | | | |
| $A$ | spheroid aspect ratio | 2.3 | | | |
| $\bar{\nu}$ | mean density within $\omega$ | 2.3 | 4 | | |
| $t$ | density ratio $\nu/\bar{\nu}$ | 2.3 | 5 | | |
| $M$ | spheroid mass | 2.3 | | | |
| $m$ | mean particle mass | 2.3 | | | |
| $k$ | spheroid mass exponent in $M \propto A^k$ | 2.3 | | | |

| Symbol | Description | Section | Eq. | Fig. |
|---|---|---|---|---|
| $Y$ | half-extent of background medium[b] | 2.4 | | |
| $N$ | column density | 2.4 | | 1,2 |
| $N_u$ | $N$ of uniform background[b] | 2.4 | | |
| $\Delta N$ | column density above background | 2.4 | 6 | 2 |
| $\Delta N_{\parallel p}$ | $\Delta N$ for parallel prolate spheroid | 2.4 | 7 | 2 |
| $\Delta N_{\perp p}$ | $\Delta N$ for perpendicular prolate spheroid | 2.4 | 8 | 2 |
| $\Delta N_{\parallel o}$ | $\Delta N$ for parallel oblate spheroid | 2.4 | 9 | 2 |
| $\Delta N_{\perp o}$ | $\Delta N$ for perpendicular oblate spheroid | 2.4 | 10 | 2 |
| $\Delta N_s$ | $\Delta N$ for $p=2$ Plummer sphere | 2.4 | | |
| $\Delta N_{\max}$ | $\Delta N$ at peak of column density map | 2.5 | | 1 |
| $x_0$ | $\Delta N$ contour ellipse horizontal radius | 2.5 | | 1 |
| $z_0$ | $\Delta N$ contour ellipse vertical radius | 2.5 | | 1 |
| $b_l$ | shortest radius of $l$-c contour[c] | 2.6 | | |
| $N_l$ | $N$ of $l$-c contour | 2.6 | 11 | |
| $M_l$ | $M$ within $l$-c contour | 2.6 | | |
| $\bar{v}_l$ | $\bar{v}$ within $l$-c contour | 2.6 | | |
| $b_u$ | short-axis radius of uniform spheroid | 2.6 | | |
| $M_u$ | $M$ within original uniform spheroid | 2.6 | | |
| $N_u$ | mean $N$ of uniform spheroid[d] | 2.6 | 12 | |
| $\epsilon_l$ | spheroid formation efficiency $M_l/M_u$ | 2.6 | | |
| $\nu_{Nl}$ | ratio $l$–c to background $N_l/N_u$ | 2.6 | 13 | |
| $\omega_l$ | normalized short-axis radius $b_l/r_0$ | 2.6 | 13 | |
| $\chi$ | scale factor $\nu_0/\omega_l^2$ | 2.6 | 14 | |
| $C$ | background parameter $\epsilon_l/(\nu_{Nl}-1)^3$ | 2.6 | | |
| $C_{\min}$ | minimum allowed value of $C$ | 2.6 | | |
| $\beta$ | normalized background ratio $C/C_{\min}$ | 2.6 | | |
| $f$ | function which relates $\beta$ to $\chi$ | 2.6 | 16 | |
| $A_{\nu\epsilon}$ | allowed area in $\nu_{Nl}$ - $\epsilon_l$ plane | 2.7.1 | | |
| $A_{\nu\epsilon,\max}$ | max allowed area in $\nu_{Nl}$ - $\epsilon_l$ plane | 2.7.1 | | |
| $\theta_B$ | $B$ field direction (W of N) in POS[e] | 2.8 | 18 | 1,4,6 |
| $s$ | coordinate ratio $\xi/\zeta$ | 2.8 | 18 | |
| $a$ | toroidal twist scale height | 2.8.3 | | |
| $\alpha$ | normalized twist scale height $a/r_0$ | 2.8.3 | 22 | |
| $t_q$ | density ratio $t$ for $\theta_B(A,i,\alpha)$ | 2.8.4 | 5, 23 | 2 |
| $\rho$ | mass density $mn$ | 3.1 | 24 | |
| $B_p$ | mean POS component of $B$ | 3.1 | 24 | |
| $\sigma_{B_p}$ | Alfvénic fluctuation amplitude for $B_p$ | 3.1 | 24 | |
| $\sigma_{NT}$ | LOS nonthermal velocity dispersion | 3.1 | 24 | |
| $Q_{B_p}$ | coefficient relating $B_p$ to DCF model | 3.1 | 24 | |
| $I$ | mean $B/B_p$ over range of $i$ | 3.1 | | |
| $\sigma_I$ | std. deviation $\sigma_B/B_p$ over range of $i$ | 3.1 | | |
| $\langle \bar{B}_1 \rangle$ | mean $B$ over inclination and within $\omega_1$ | 3.1 | 25 | |
| $Q_B$ | coefficient relating $B$ to DCF model | 3.1 | 25 | |

| Symbol | Description | Section | Eq. | Fig. | Table |
|---|---|---|---|---|---|
| $\sigma_B$ | Alfvénic fluctuation amplitude for $B$ | 3.1 | 25 | | |
| $\bar{\rho}_1$ | mean $\rho$ within $\omega_1$ | 3.1 | 25 | | |
| $(\bar{\sigma}_{NT})_1$ | mean $\sigma_{NT}$ within $\omega_1$ | 3.1 | 25 | | |
| $\sigma_\theta$ | dispersion of observed pol angles $\theta_{B,\text{pol}}$ | 3.1 | | | |
| $\Delta\theta$ | angle difference $\theta_{B,\text{pol}}$ - model $\theta_B$ | 3.1 | | | 5,7 |
| $\sigma_{\Delta\theta}$ | dispersion of $\Delta\theta$ | 3.1 | 26 | | 5,7 |
| $M_c$ | magnetic critical mass | 3.2 | 27 | | |
| $G$ | gravitational constant | 3.2 | 27 | | |
| $\Phi$ | magnetic flux | 3.2 | | | |
| $c_\Phi$ | critical mass coefficient $M_c G^{1/2} \Phi^{-1}$ | 3.2 | 27 | | |
| $B_u$ | $B$ in original uniform medium | 3.3 | 29 | | |
| $B_0$ | maximum field strength | 3.3 | 31 | | |
| $S_\nu(v)$ | spectral line flux density | 5.1 | | | |
| $W(\text{C}^{18}\text{O})$ | integrated intensity $J = 2 - 1$ $\text{C}^{18}\text{O}$ line | 5.1 | | | 2,4 |
| $[\text{C}^{18}\text{O}]$ | abundance $N(\text{C}^{18}\text{O})/N$ | 5.1 | | | |
| $\widetilde{W}$ | $W(\text{C}^{18}\text{O})$ normalized by rms noise | 5.1 | 32 | | |
| $\widetilde{N}$ | $N$ normalized by $\Delta N_{\max}$ | 5.1 | | | |
| $C_N$ | scale factor $\widetilde{W}/\widetilde{N}$ | 5.1 | 32 | | 2 |
| $\sigma_b$ | $e^{-1}$ radius of Gaussian beam | 5.1 | | | |
| $r_{0s}$ | fit scale length smoothed by beam | 5.1 | | | |
| $\sigma_n$ | uncertainty in density $n$ | 5.2 | | | |
| $r_{\text{scat}}$ | smallest radius where SFF model applies | 5.3.2 | | | 5 |
| $r_{\max}$ | largest radius where SFF model applies | 5.3.2 | | | 5 |
| $r_{\max,0}$ | value of $r_{\max}$ which gives smallest $\sigma_{\Delta\theta}$ | 5.3.2 | | | 5 |

**Notes.** Columns (3) and (4) list the section and equation number where each symbol is first used. Columns (5) and (6) list a relevant figure and table number where the symbol also appears.

[a]LOS = line of sight; [b]in simple background model; [c]l – c contour = lowest-closed column density contour; [d]in detailed background model; [e]POS = plane of sky